\documentclass{emulateapj}

\usepackage{natbib}

\usepackage{graphicx}
\usepackage{epstopdf}
\usepackage{amsmath}
\usepackage{amssymb}
\usepackage{bm}
\usepackage{color}

\def\gs{\mathrel{\raise0.35ex\hbox{$\scriptstyle >$}\kern-0.6em
\lower0.40ex\hbox{{$\scriptstyle \sim$}}}}
\def\ls{\mathrel{\raise0.35ex\hbox{$\scriptstyle <$}\kern-0.6em
\lower0.40ex\hbox{{$\scriptstyle \sim$}}}}

\newcommand{\msun}{$\rm{M}_{\odot}$}

\newcommand{\mbh}{M$_{\rm BH}$}

\newcommand{\Hb}{H$\beta$}

\newcommand{\msigma}{M$_{\rm BH}-\sigma_*$}

\newcommand{\heii}{He\,{\sc ii}}
\newcommand{\hei}{He\,{\sc i}}
\newcommand{\feii}{Fe\,{\sc ii}}
\newcommand{\mgii}{Mg\,{\sc ii}}
\newcommand{\civ}{C\,{\sc iv}}
\newcommand{\oiii}{[O\,{\sc iii}]}

\input{commands.input}

\shorttitle{The Broad Line Region of Arp 151} 
\shortauthors{Pancoast et al.}
 
\begin{document}

\title{Stability of the Broad Line Region Geometry and Dynamics \\
in Arp 151 Over Seven Years} 

\author{
 A.~Pancoast\altaffilmark{1,2}, 
 A.~J.~Barth\altaffilmark{3}, 
 K.~Horne\altaffilmark{4},
 T.~Treu\altaffilmark{5},
 B.~J.~Brewer\altaffilmark{6},
 V.~N.~Bennert\altaffilmark{7},
 G.~Canalizo\altaffilmark{8}, 
 E.~L.~Gates\altaffilmark{9},
 W.~Li\altaffilmark{10, 22},
  M.~A.~Malkan\altaffilmark{5},
 D.~Sand\altaffilmark{11},
 T.~Schmidt\altaffilmark{3},
  S.~Valenti\altaffilmark{12},
 J.~H.~Woo\altaffilmark{13},
 K.~I.~Clubb\altaffilmark{10}, 
 M.~C.~Cooper\altaffilmark{3},
 S.~M.~Crawford\altaffilmark{14},
 S.~F.~H\"{o}nig\altaffilmark{15},
 M.~D.~Joner\altaffilmark{16}, 
 M.~T.~Kandrashoff\altaffilmark{10},
 M.~Lazarova\altaffilmark{17},
 A.~M.~Nierenberg\altaffilmark{3}, 
 E.~Romero-Colmenero\altaffilmark{14, 18},
 D.~Son\altaffilmark{13},
 E.~Tollerud\altaffilmark{19}, 
 J.~L.~Walsh\altaffilmark{20}, and
 H.~Winkler\altaffilmark{21}
}

% Pancoast
 \altaffiltext{1}{Harvard-Smithsonian Center for Astrophysics, 60 Garden Street, Cambridge, MA 02138, USA; anna.pancoast@cfa.harvard.edu}

% Pancoast
 \altaffiltext{2}{Einstein Fellow}

% Barth, Nierenberg, Cooper
 \altaffiltext{3}{Department of Physics and Astronomy, 4129 Frederick Reines Hall, University of California, Irvine, CA, 92697-4575, USA}
 
  % Horne
 \altaffiltext{4}{SUPA Physics and Astronomy, University of St.  Andrews, Fife, KY16 9SS Scotland, UK}

 % Treu
 \altaffiltext{5}{Department of Physics and Astronomy, University of California, Los Angeles, CA 90095-1547, USA}

% Brewer
 \altaffiltext{6}{Department of Statistics, The University of Auckland, Private Bag 92019, Auckland 1142, New Zealand}
 
 % Bennert
 \altaffiltext{7}{Physics Department, California Polytechnic State University, San Luis Obispo, CA 93407, USA}
 
 % Canalizo
\altaffiltext{8}{Department of Physics and Astronomy, University of California, Riverside, CA 92521, USA}
 
 % Gates
 \altaffiltext{9}{Lick Observatory, P.O. Box 85, Mount Hamilton, CA 95140, USA}
 
 % Li, Clubb, Kandrashoff
\altaffiltext{10}{Department of Astronomy, University of California, Berkeley, CA 94720-3411, USA} 
 
 % Sand
 \altaffiltext{11}{Department of Astronomy and Steward Observatory, University of Arizona, 933 N Cherry Ave, Tucson, AZ 85719, USA}
 
% Valenti
 \altaffiltext{12}{Las Cumbres Observatory Global Telescope Network, 6740 Cortona Drive, Suite 102, Goleta, CA 93117, USA}
 
 % Woo, Son
 \altaffiltext{13}{Astronomy Program, Department of Physics and Astronomy, Seoul National University, Seoul 151-742, Korea}

% Crawford, Romero-Colmenero
\altaffiltext{14}{South African Astronomical Observatory, P.O. Box 9, Observatory 7935, Cape Town, South Africa}

% Honig
\altaffiltext{15}{School of Physics \& Astronomy, University of Southampton, Southampton SO17 1BJ, UK}

% Joner
\altaffiltext{16}{Department of Physics and Astronomy, N283 ESC, Brigham Young University, Provo, UT 84602-4360, USA}

% Lazarova
\altaffiltext{17}{Dept.of Physics and Astronomy, University of Nebraska at Kearney, Kearney, NE 68849, USA}

% Romero-Colmenero
\altaffiltext{18}{Southern African Large Telescope Foundation, P.O. Box 9, Observatory 7935, Cape Town, South Africa}

% Tollerud
\altaffiltext{19}{Space Telescope Science Institute, 3700 San Martin Dr., Baltimore, MD 21218, USA}

% Walsh
\altaffiltext{20}{George P. and Cynthia Woods Mitchell Institute for  Fundamental Physics and Astronomy, Department of Physics and  Astronomy, Texas A\&M University, College Station, TX 77843-4242,  USA}

% Winkler
\altaffiltext{21}{Department of Physics, University of Johannesburg, P.O. Box 524, 2006 Auckland Park, South Africa}

\altaffiltext{22}{Deceased 2011 December 12.}

\begin{abstract}
The Seyfert 1 galaxy Arp 151 was monitored as part of three reverberation mapping 
campaigns spanning $2008-2015$. 
We present modeling of these velocity-resolved reverberation mapping 
datasets using a geometric and dynamical model for the broad line region (BLR).  By modeling each of 
the three datasets independently, we infer the evolution of the BLR structure in Arp 151 
over a total of seven years and constrain the systematic uncertainties in non-varying parameters such 
as the black hole mass.  We find that the BLR geometry of a thick disk viewed close to face-on is stable 
over this time, although the size of the BLR grows by a factor of $\sim 2$.
The dynamics of the BLR are dominated by inflow and the inferred black hole mass is
consistent for the three datasets, despite the increase in BLR size.  
Combining the inference for the three datasets yields a black 
hole mass and statistical uncertainty of $\log_{10}($\mbh$/$\msun$)=$\,\totMbh\ with a standard deviation in individual
measurements of $0.13$ dex.  
\end{abstract}

\keywords{galaxies: active -- galaxies: nuclei -- galaxies: Seyfert -- galaxies: individual (Arp 151)}

%%%%%%%%%%%%%%%%%%%%%%%%%%%%%%%%%%%%%%
%%%%%%%%%%%%%%%%%%%%%%%%%%%%%%%%%%%%%%
%%%%%%%%%%%%%%%%%%%%%%%%%%%%%%%%%%%%%%
%%%%%%%%%%%%%%%%%%%%%%%%%%%%%%%%%%%%%%

\section{Introduction}   
\label{sect_intro}
Outside the local Universe,  
the most promising method to measure the masses of supermassive black holes (BHs) is the reverberation mapping technique \citep{blandford82, peterson93, peterson04} in accreting BHs in active galactic nuclei (AGN).   Unlike dynamical BH mass measurement techniques that require spatially resolving the BH gravitational sphere of influence and thus are limited to local and larger BHs within $\sim 150$ Mpc \citep[e.g.][]{mcconnell13}, reverberation mapping resolves the motions of gas around the BH temporally and is now being applied to AGN at redshifts of $z > 2$ \citep{kaspi07, king15, shen15a}.   
Reverberation mapping relies on the variability of AGN continuum emission from the BH accretion disk as it is reprocessed by gas in the broad emission line region (BLR).  By monitoring an AGN with spectroscopy covering one or more broad emission lines and photometry of the AGN accretion disk continuum emission, we can measure a time lag $\tau$ between variations seen in the continuum and later seen in the broad emission lines that is used as a size estimate of the BLR.  Combining the time lag with the velocity of the BLR gas $v$, as measured from the width of the broad emission lines, we can measure the black hole mass:
\begin{equation}
M_{\rm BH} = f v^2 c \tau/G
\end{equation}
where $c$ is the speed of light, $G$ is the gravitational constant, and $f$ is a factor of order unity that depends on the geometry, dynamics, and orientation of the BLR.  While $\tau$ and $v$ are often measured to better than 20\%, the value of $f$ is generally unknown in individual AGN because the BLR is not spatially resolved.  The unknown value of $f$ thus introduces the largest source of uncertainty in reverberation mapped BH masses and requires the use of an average value that is calibrated by assuming the same \mbh$-\sigma_*$ relation for active and inactive galaxies
\citep{onken04, collin06, woo10, greene10b, graham11, park12b, woo13, grier13b, woo15, batiste17}.  While an average value for $f$ has allowed reverberation mapped BH masses to be measured in over 50 AGN \citep[for a compilation see][]{bentz15}, it introduces an uncertainty in individual BH masses that could be as large as the scatter in the \mbh$-\sigma_*$ relation of $\sim 0.4$ dex \citep[a factor of $\sim 2.5$; see e.g.][]{park12b}.  The only way to reduce this uncertainty is to understand the detailed structure of the BLR and hence measure the value of $f$ in individual AGN.  

The drive to better understand the BLR combined with the small sample size of reverberation mapped AGN has motived two complimentary approaches. 
The first approach is to substantially increase the reverberation mapping sample in terms of both the number of AGN and of which broad emission lines are used \citep{king15, shen15a}, with \Hb\ in the local Universe being replaced by \mgii\,$\lambda$2799 and \civ\,$\lambda$1549  as the broad emission lines of choice at higher redshift.
While reverberation mapping becomes more challenging at higher redshifts due to longer time lags and lower AGN variability \citep[e.g.][]{macleod10}, the tight relation between the BLR size and AGN luminosity for \Hb, the $r_{\rm BLR}-L_{\rm AGN}$ relation \citep{kaspi00, kaspi05, bentz09a, bentz13}, allows for thousands of single-epoch BH mass estimates \citep[e.g.][]{shen11} made using a single spectrum in place of long-term spectroscopic and photometric monitoring \citep[e.g.][and references therein]{vestergaard06, shen13, park17}.  One of the main goals of larger reverberation mapping campaigns monitoring \mgii\ and \civ\ lines is to reduce the uncertainty in single-epoch masses using these lines by measuring their $r_{\rm BLR}-L_{\rm AGN}$ relations directly instead of calibrating BH masses relative to \Hb\ \citep[e.g.][]{shen16}.  However, the unknown structure of the BLR still introduces a large uncertainty in estimates of the BH mass, since different broad emission lines probe different regions of the BLR due to ionization stratification \citep{clavel91, reichert94} and some lines may be more sensitive to non-gravitational forces from AGN winds.  

The second approach, of generating a few high-quality reverberation mapping datasets with mainly \Hb\ in the local Universe, is therefore critical to understanding single-epoch BH masses as well, since it provides detailed information about the geometry, dynamics, and orientation of the BLR.  In the highest-quality datasets, the time lag can be measured in velocity (or wavelength) bins across the broad emission line and the structure of these velocity-resolved lags is interpreted in terms of the general dynamics of the BLR.  Such velocity-resolved reverberation mapping datasets are still in the minority, but the recent application of detailed analysis techniques has uncovered a wealth of information and the first direct measurements of $f$ in individual AGN \citep[e.g.][]{pancoast14b}.  Some of the most successful velocity-resolved reverberation mapping datasets include the MDM 2007 \citep{denney10}, 2010 \citep{grier12}, 2012, and 2014 \citep{fausnaugh17} campaigns, the Lick AGN Monitoring Project (LAMP) 2008 \citep{walsh09, bentz09} and 2011 \citep{barth15} campaigns, the Lijiang 2012-2013 \citep{du16} and 2015 \citep{lu16} campaigns, and the AGN STORM 2014 campaign \citep{derosa15, edelson15, fausnaugh16, pei17}.  
 
The ultimate goal of these campaigns is to understand the velocity-resolved response of the broad line flux $L(v_{\rm LOS}, t)$ at line-of-sight velocity $v_{\rm LOS}$ and observed time $t$ to the AGN continuum variability $C$:
\begin{equation}
\label{eq_2}
L(v_{\rm LOS}, t) = \int^\infty_0 \Psi (v_{\rm LOS}, \tau) C(t-\tau) d\tau,
\end{equation}
where $C(t-\tau)$ is the continuum flux at earlier time $t-\tau$ and $\Psi$ is the transfer function that relates the line and continuum emission as a function of $v_{\rm LOS}$ and time lag $\tau$ \citep{blandford82}.  When changes in the line and continuum fluxes from their mean values are used in Equation~\ref{eq_2} instead of total flux values,  $\Psi$ is called the response function and is generally not the same as the transfer function \citep[for a discussion of this difference see Section 4 of][]{goad15}.
Detailed analysis of velocity-resolved reverberation mapping datasets has focused on either constraining the velocity-resolved response function in a model independent context or on modeling the reverberation mapping data with a BLR model directly.  Both approaches have their merits: constraining the response function requires fewer assumptions about BLR physics, while modeling the BLR directly requires more assumptions about the physics but yields quantitative constraints on model parameter values.  Response functions have been measured using the MEMEcho code \citep{horne91, horne94} for the LAMP 2008 dataset for Arp 151 \citep{bentz10} and the 2010 MDM campaign \citep{grier13a}, while response functions have been measured using regularized linear inversion \citep{krolik95, done96} for LAMP 2008 \citep{skielboe15}.  Modeling reverberation mapping data directly using a geometric and dynamical BLR model has been done for the LAMP 2008 dataset for Arp 151 \citep{brewer11}, the LAMP 2011 dataset for Mrk 50 \citep{pancoast12}, five AGN from the LAMP 2008 dataset \citep{pancoast14b}, and four AGN from the MDM 2010 dataset \citep{grier17}.  In addition, a geometry-only BLR model has been applied to a larger AGN sample that includes the LAMP 2008 and MDM 2010 samples \citep{li13}.  Both approaches of constraining the response function and modeling the BLR show that the BLR dynamics can vary widely between AGN. 
Response functions can show symmetry of the red and blue sides of the line, interpreted as gas orbiting in a disk.
Alternatively, response functions can show response at longer time delays on either 
the blue \citep[e.g. Arp 151;][]{bentz10} 
or red side of the emission line \citep[e.g. the velocity-resolved lag measurements for Mrk 3227;][]{denney09c} 
usually interpreted as signatures of inflowing or outflowing gas, respectively \citep[although see][for prompt red-side response in the context of an outflow model]{bottorff97}.  Modeling the BLR directly shows similar features as MEMEcho in the resulting transfer functions.

However, there are still many unanswered questions about the structure of the BLR that could significantly impact our ability to measure BH masses with reverberation mapping techniques.  One unknown is the extent to which BLR structure evolves over time.  As shown by both the $r_{\rm BLR}-L_{\rm AGN}$ relation for the reverberation mapped sample and for individual sources such as NGC 5548 \citep{pei17}, the size of the BLR can change significantly in response to variability in the AGN continuum.  Substantial changes in the AGN accretion rate could affect at least the dynamics of the emitting gas through the generation of AGN-driven outflows \citep[e.g.][]{emmering92, murray97, proga00, elvis00}.  While values of the virial product $\tau v^2$ measured for the same source over time have shown that the BH mass is consistently measured for different broad emission lines and values of the time lag $\tau$ \citep[e.g.][]{peterson99, peterson00, pei17}, the uncertainties are still large and it is possible that the virial factor $f$ changes with time as well.  Another unknown is the full uncertainty with which BLR properties are currently being measured through the forward-modeling approaches of \citet{pancoast14a} and \citet{li13}.  Since BLR modeling is necessary in order to quantify information directly from reverberation mapping data or through the inferred response function, having a complete understanding of both statistical and systematic uncertainties is needed in order to completely rule out classes of models.  Systematic uncertainties can be introduced at every stage of the BLR modeling approach, from how the reverberation mapping data are pre-processed to isolate the emission line flux to the choice of model parameterization and what physics is included for the AGN continuum source and BLR gas.

In this paper we aim to address these questions by applying the BLR modeling approach to three velocity-resolved reverberation mapping datasets for Arp 151 (also called Mrk 40) taken over seven years, corresponding to the orbital time for gas $\sim 4$ light days from the BH.  The main goals of this analysis are to 1) probe possible evolution in BLR structure, 2) investigate the reproducibility of BLR modeling results for independent datasets of the same source, and 3) constrain possible sources of systematic uncertainty in BLR modeling analysis.  
The three datasets are described in Section~\ref{sect_data} and a brief overview of the BLR model is given in Section~\ref{sect_model}.  Results from BLR modeling analysis of each of the datasets individually as well as a joint inference on BLR structure are detailed in Section~\ref{sect_results}.  
Possible evolution in the BLR geometry and dynamics and the effects of systemic uncertainties are described in Section~\ref{sect_discussion}, along with a comparison between the transfer functions from BLR modeling and the response functions from MEMEcho analysis.
Finally, a summary of our work is given in Section~\ref{sect_summary}.
All final values for distances and BH masses from modeling the BLR are given in the rest frame of the AGN.  To convert to the observed frame, multiply distances and the black hole mass by $1+z = 1.021091$, where $z$ is the redshift.

%%%%%%%%%%%%%%%%%%%%%%%%%%%%%%%%%%%%%%
%%%%%%%%%%%%%%%%%%%%%%%%%%%%%%%%%%%%%%
%%%%%%%%%%%%%%%%%%%%%%%%%%%%%%%%%%%%%%
%%%%%%%%%%%%%%%%%%%%%%%%%%%%%%%%%%%%%%

\section{Data}   \label{sect_data}
We now describe the three velocity-resolved reverberation mapping datasets for Arp 151
with sampling, resolution, and data quality characteristics listed in Table~\ref{table_data}.

 \begin{deluxetable*}{lccc}
%\tabletypesize{\scriptsize}
\tablecaption{Data Characteristics}
\tablewidth{0pt}
\tablehead{ 
\colhead{$\,\,\,\,\,\,\,\,\,\,\,\,$} & 
\colhead{LAMP 2008} &
\colhead{LAMP 2011} &
\colhead{LCO 2015} 
}
\startdata
(1) Dates    &  March 25 - June 1, 2008  & March 27 - June 13, 2011 & December 6, 2014 - June 5, 2015 \\
(2) $n_{\rm cont}$     &  84       & 91       & 119         \\
(3) $n_{\rm line}$     &   43       & 39       & 55          \\
(4) $\Delta t_{\rm cont}$ (days)     &   0.93       & 0.94       & 1.03        \\  
(5) $\Delta t_{\rm line}$ (days)    &   1.02       & 1.04       & 1.51     \\
(6) Spectral S/N  &   80       & 72       & 22.5     \\
(7) $\Delta \lambda_{\rm instru}$ (\AA)   & 5.06     & 2.47    & 3.44     \\
(8) Pixel scale (\AA)    & 2    & 1     & 1.74    \\
(9) Slit width                &  4\arcsec   &  4\arcsec         &  $1\arcsec.6 $  \\
(10) Extraction width   &  10\arcsec.1   &  10\arcsec.3    &    $8\arcsec.8$  
\enddata
\tablecomments{The table rows are as follows: (1) the range of dates for spectroscopic monitoring, 
(2) the number of epochs in the AGN continuum light curve, (3) the number of epochs of spectroscopy,
(4) median time between continuum epochs, 
(5) median time between spectral epochs,
(6) median S/N ratio of the spectra in the optical continuum,
(7) instrumental resolution \citep[line dispersion, $\sigma$; see Sect. 2 of][]{pancoast14b}, 
(8) spectral pixel scale,
(9) slit width of spectroscopy, and
(10) extraction region width used to generate spectra.
Values for items (6), (9), and (10) are taken from \citet{bentz09}, \citet{barth15}, and \citet{valenti15} for LAMP 2008, LAMP 2011, and LCO 2015, respectively.
  \label{table_data}}
\end{deluxetable*} 

\subsection{LAMP 2008}
The LAMP 2008 reverberation mapping campaign was the first to target Arp 151 and
includes photometric monitoring of the AGN continuum in the {\it B} and {\it V} bands  \citep{walsh09}
and spectroscopic monitoring in the optical from $4300-7100$~\AA\ \citep{bentz09}.  
The AGN continuum light curves for Arp 151
were measured with standard aperture photometry techniques using data from the
0.80 m Tenagra II telescope in southern Arizona that is part of the Tenagra 
Observatories complex. 
Optical spectroscopy is from the Kast Spectrograph \citep{miller93} 
on the 3 m Shane Telescope at Lick Observatory.  

\subsection{LAMP 2011}
Arp 151 was monitored again as part of the LAMP 2011 reverberation mapping campaign,
with photometric monitoring in the {\it V} band (Pancoast et al., in preparation) and spectroscopic monitoring in the 
optical from $3440-8200$~\AA\ \citep{barth15}.  The AGN continuum light curve for Arp 151
was measured with difference imaging techniques using data from multiple telescopes, 
including the 0.91 m telescope at West Mountain Observatory, the 2 m Faulkes Telescopes 
North and South in the Las Cumbres Observatory network \citep[LCO;][]{brown13}, the 0.76 m 
Katzman Automatic Imaging Telescope at Lick Observatory \citep{filippenko01}, and the 0.6 m Super-LOTIS 
telescope at Steward Observatory, Kitt Peak.
Optical spectroscopy is from the Kast Spectrograph \citep{miller93}
on the 3 m Shane Telescope at Lick Observatory. 

\subsection{LCO AGN Key Project 2015}
Arp 151 was also monitored as part of the ongoing LCO AGN Key Project reverberation 
mapping campaign (hereafter LCO 2015) with photometric monitoring of the AGN continuum in the {\it V} band
and spectroscopic monitoring in the optical from $\sim3200-10000$~\AA\ \citep{valenti15}.  
The AGN continuum light curve for Arp 151
was measured with the automated aperture photometry scripts described by \citet{pei14} 
using data from LCO network telescopes, 
including the 1 m telescope at McDonald Observatory and the 2 m Faulkes Telescope North.   
Optical spectroscopy is from the FLOYDS Spectrograph 
on the 2 m Faulkes Telescope North.   

\subsection{Spectral Decomposition}

\begin{figure*}[h!]
\begin{center}
\includegraphics[scale=0.55]{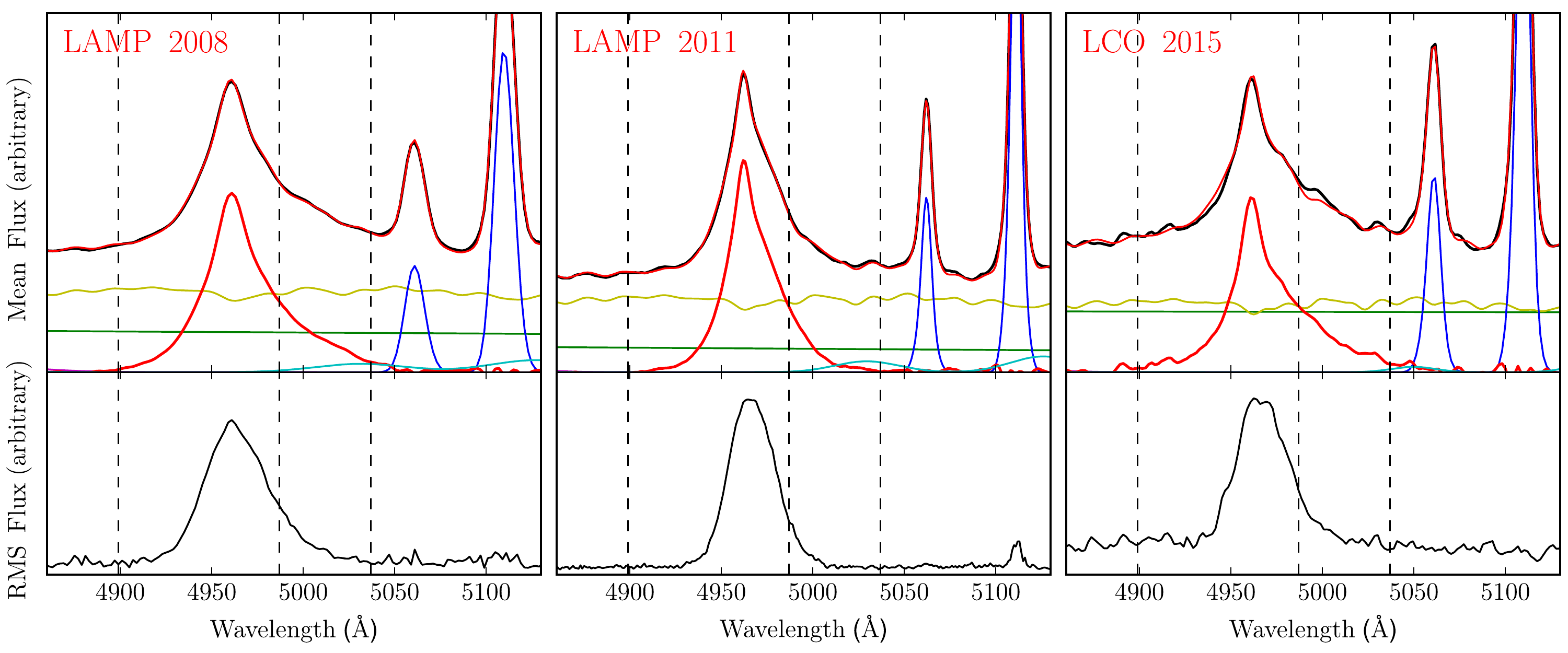}  % left, bottom, right, top
\caption{Spectral decomposition of the mean spectrum (top panels) and the RMS line profile for the \Hb\ component (bottom panels) for the LAMP 2008, LAMP 2011, and LCO 2015 Arp 151 datasets.  The vertical dashed lines show the wavelength ranges used for modeling the BLR.  In addition to the mean data spectrum (black line), the spectral decomposition components shown in the top panel include the full model (top red line), AGN power-law continuum (green line), host galaxy starlight (yellow line), narrow \oiii\ emission lines (blue line), \heii\ (magenta lines, mostly too faint at the wavelengths shown), \feii\ (cyan line), and \Hb\ component constructed by subtracting all components except for broad and narrow \Hb\ from the data (bottom red line).   The wavelengths shown on the x-axis are the observed wavelengths and the flux units for both the mean and RMS spectra are in arbitrary units.   \label{fig_spect_decomp}}
\end{center}
\end{figure*}

The H$\beta$ broad and narrow emission lines used in our analysis were isolated from
the optical spectrum using the spectral decomposition scripts described by \citet{barth13}.
The spectra were decomposed into contributions from the power-law AGN continuum, the host galaxy starlight,
emission lines for \Hb, \heii, and \oiii, and a template for \feii\ emission blends.
While the spectral decomposition scripts allow for additional components of 
\hei\ at 4471, 4922, and 5016~\AA,
these components were not included in the fit because they were not consistently differentiated from \feii\ and other overlapping features in
the red wing of \Hb\ for individual spectral epochs.  
The blue wing of \Hb\ is less contaminated, with no substantial overlap with \heii. 
An example of the spectral
decomposition in the region around \Hb\ is shown for each dataset in Figure~\ref{fig_spect_decomp} along
with the root-mean-square (RMS) spectrum.  Note that the RMS flux in the continuum is higher for the LCO 2015 dataset 
due to the lower signal-to-noise (S/N) ratio of the spectra.   

The spectral decomposition was done using three different \feii\ templates from
\citet{boroson92}, \citet{veron-cetty04}, and \citet{kovacevic10}.  While the results
from the three \feii\ templates are often very similar \citep[e.g.][]{barth15}, sometimes there are differences
in the integrated emission line light curve scatter or in the root-mean-square (RMS) 
spectrum.  For our analysis of Arp 151, we use the \citet{kovacevic10} template because 
it is able to better fit the data.  Both the $\chi^2$ and the
reduced $\chi^2$, which compensates for the larger number of free parameters, are
smallest for the \citet{kovacevic10} template.  The integrated emission line light curve scatter 
and the RMS spectrum are very similar for the different templates.

To reduce systematic uncertainties introduced by assuming a smooth model for \Hb\
in the spectral decomposition, we isolate the \Hb\ emission to be used in our
analysis by subtracting all spectral decomposition components from the data
except for broad and narrow \Hb.  Examples of the isolated \Hb\ emission are also
shown in Figure~\ref{fig_spect_decomp}.  While the spectral decomposition script
does not provide statistical uncertainties from the spectral modeling, tests using a Monte-Carlo procedure 
to estimate the final flux uncertainties for LAMP 2011 suggest that the additional statistical
uncertainty introduced by spectral decomposition is very small \citep{barth15}.  We therefore use
the original statistical uncertainties of the optical spectra for our analysis.

%%%%%%%%%%%%%%%%%%%%%%%%%%%%%%%%%%%%%%
%%%%%%%%%%%%%%%%%%%%%%%%%%%%%%%%%%%%%%
%%%%%%%%%%%%%%%%%%%%%%%%%%%%%%%%%%%%%%
%%%%%%%%%%%%%%%%%%%%%%%%%%%%%%%%%%%%%%

\section{The Geometric and Dynamical Model of the Broad Line Region}  \label{sect_model}
In this Section we give an overview and define the model parameters of our 
parameterized, phenomenological model for the BLR.  A full description is given 
by \citet{pancoast14a}.  

We model the distribution of broad line emission using many massless point test particles
that linearly and instantaneously reprocess the AGN continuum flux from the accretion
disk into emission line flux seen by the observer.  The accretion disk is assumed
to be a point source at the origin that emits isotropically.  The accretion disk photons
are reprocessed into line emission with a time lag that is determined by a point particle's 
position and a wavelength of emitted line flux that is determined by a point particle's velocity. 

In order to calculate the line emission from each point particle at any given time,
we need to know the AGN continuum flux between the data points in the light curve. 
We generate a continuous model of the AGN continuum light curve using 
Gaussian processes.  By simultaneously exploring the parameter space of the 
Gaussian process model parameters, we can include the uncertainties from 
interpolation between the AGN continuum light curve data points into our 
inference of the BLR model parameters.  We can also use the AGN continuum
light curve model to extrapolate to earlier or later times beyond the extent
of the data in order to evaluate the contribution of point particles with long time lags 
at the beginning of the campaign and to model the response of the BLR after
the AGN continuum monitoring has ended.  

\subsection{Geometry}
The radial distribution of point particles is parameterized by a
Gamma distribution:
\begin{eqnarray}
p(r|\alpha, \theta)  \propto r^{\alpha-1} \exp \left( - \frac{r}{\theta}   \right)
\end{eqnarray}
that is shifted radially from the origin by the Schwarzschild radius, $R_s = 2GM_{\rm BH}/c^2$, 
plus a minimum radius of the BLR, $r_{\rm min}$.  This shifted Gamma distribution is 
also truncated at an outer radius $r_{\rm max}$ (listed in Table~\ref{table_results}).   
We perform a change of variables from $(\alpha, \theta, r_{\rm min})$ to 
$(\mu, \beta, F)$ in order to work in units of the mean radius, $\mu$, such that:
\begin{eqnarray}
\mu &=& r_{\rm min} + \alpha \theta \\
\beta &=& \frac{1}{\sqrt{\alpha}} \\
F &=&  \frac{r_{\rm min}}{r_{\rm min} + \alpha \theta}
\end{eqnarray}
where $\beta$ is the shape parameter and $F$ is $r_{\rm min}$ in units of $\mu$.  
The standard deviation of the radial distribution is then given by $\sigma_r = (1-F)\mu \beta$.  
For the three free parameters, $(\mu, \beta, F)$, the prior probability distribution 
is uniform in the log of the parameter between $1.02 \times 10^{-3}$ light days and $r_{\rm out}$ 
for $\mu$, uniform between 0 and 2 for $\beta$, and uniform between 0 and 1 for $F$.

Spherical symmetry is broken by defining a half-opening angle of the point particles, $\theta_o$, such
that values of $\theta_o \to 0$ (90) degrees correspond to a thin disk (spherical) geometry with a uniform 
prior between 0 and 90 degrees.
An observer views the BLR from an inclination angle, $\theta_i$, where $\theta_i \to 0$ (90) degrees 
corresponds to a face-on (edge-on) orientation, and the prior is uniform in $\cos (\theta_i)$ between 0 and 90 
degrees.
The emission from each point particle is given a relative weight, $W$, between 0 and 1:
\begin{eqnarray}
W(\phi) = \frac{1}{2} + \kappa \cos(\phi)
\end{eqnarray}
where $\kappa$ is a free parameter with a uniform prior between $-0.5$ and 0.5 and $\phi$ is the angle 
between the observer's line of sight to the origin and the point particle's line of sight to the origin.  
When $\kappa \to -0.5$ (0.5) then the far (near) side of the BLR is contributing more line emission.
We also allow the point particles to be clustered near the faces of the disk, such
that the angle $\theta$ of a point particle from the disk is:
\begin{eqnarray}
\theta = {\rm arccos} (\cos \theta_o + (1- \cos \theta_o) U^{\gamma} )
\end{eqnarray}
where $U$ is a random number drawn uniformly between 0 and 1 
and $\gamma$ is a free parameter
with a uniform prior between 1 and 5.  When $\gamma \to 1$ (5), point 
particles are evenly distributed in (clustered at the faces of) the disk.  
Finally, we allow for a transparent to opaque disk mid-plane, where
$\xi$ is twice the fraction of point particles below the disk mid-plane. 
When $\xi \to 0$ (1), the mid-plane 
is opaque (transparent) with a uniform prior between 0 and 1.

\subsection{Dynamics}
The velocities of the point particles depend upon the black hole mass, $M_{\rm BH}$, which has
a uniform prior in the log of the parameter between $2.78 \times 10^4$ and $1.67 \times 10^9 \,M_\odot$. 
We define two types of Keplerian orbits for the point particles in the plane of their radial and tangential velocities.  
The first type of orbit is drawn from a Gaussian distribution in the 
radial and tangential velocity plane centered on the circular orbit value, resulting in
bound, elliptical orbits.  A fraction, $f_{\rm ellip}$, of the point particles have these near-circular elliptical orbits, 
where $f_{\rm ellip}$ has a uniform prior between 0 and 1 and $f_{\rm ellip} \to 0$ (1) corresponds to no (all)
point particles with velocities of this type.
  
The remaining point particles have a second type of orbit that 
is drawn from a Gaussian distribution centered
on the radial inflowing or outflowing escape velocity, where values of 
$0<f_{\rm flow} < 0.5$ designate inflow and $0.5<f_{\rm flow} < 1$
designate outflow and $f_{\rm flow}$ has a uniform prior between 0 and 1.
The center of the distribution of this second type of orbit can also
be rotated on an ellipse towards the circular orbit value by an angle $\theta_e$
that has a uniform prior between 0 and 90 degrees. 
Finally, for each point particle we include a contribution from randomly-oriented macroturbulent velocities
with magnitude:
\begin{eqnarray}
v_{\rm turb} = \mathcal{N}(0, \sigma_{\rm turb})|v_{\rm circ}|
\end{eqnarray}
where $v_{\rm circ}$ is the circular velocity at the point particle's radius 
and $\mathcal{N}(0, \sigma_{\rm turb})$ is a random
number drawn from a Gaussian distribution with a mean of 0 and a standard deviation of $\sigma_{\rm turb}$.  
The prior of $\sigma_{\rm turb}$ is uniform in the log of the parameter between 0.001 and 0.1.

\subsection{Generating Model Spectra}
For a specific set of model parameter values the positions, velocities, and weights of each 
point particle are determined, and using a Gaussian process model for the AGN 
continuum light curve we can generate a time-series of model emission line 
profiles in velocity space.
To convert the model spectra to wavelength space we include 
the effects of relativistic doppler shift and gravitational redshift.
The model spectra are also blurred by the time-variable instrumental resolution
of the spectroscopic monitoring data, which is measured by comparing the 
intrinsic width of the narrow \oiii\,$\lambda 5007$ emission line measured by \citet{whittle92} to the measured
width for each observation, with typical values listed in Table~\ref{table_data}.  We also model the narrow component 
of the \Hb\ emission line as a Gaussian with the same intrinsic width as
\oiii\,$\lambda 5007$ blurred by the instrumental resolution.  
Finally, due to the importance of defining the center of 
the broad emission line and thus the region where point particles with zero
line-of-sight velocity contribute line flux, the 
systematic central wavelength is a free parameter
with a narrow Gaussian prior with standard deviation of $1$~\AA\ for LAMP 2008 and 
$4$~\AA\ for LAMP 2011 and LCO 2015.  The implications of this choice for the 
standard deviation of the 
systematic central wavelength 
prior are discussed in Section~\ref{sect_simple_model}.   

\subsection{Exploring the Model Parameter Space}
We explore the high-dimensional parameter space of the AGN continuum light curve model and
the BLR model using the diffusive nested sampling code DNest3 \citep{brewer11}. 
Diffusive nested sampling provides posterior probability distribution functions (PDFs)
and calculates the ``evidence", allowing for comparison of models that are parameterized 
differently.  
We compare the broad and narrow emission line models to the broad and narrow
emission line data using a Gaussian likelihood function.  
In order to calculate posterior PDFs in post-processing, 
we soften the likelihood function by dividing the log of the likelihood
by a temperature $T$, where $T \ge 1$.  Using values of $T>1$ accounts
for effects such as underestimated uncertainties in the spectral data or
the inability of a simple model to fit the full complexity of the data.  For 
a Gaussian likelihood function, setting $T>1$ is equivalent to increasing
the uncertainties on the spectral data by $\sqrt{T}$.  
We use values of $T = 65$ (45) for LAMP 2008, 60 (45) for LAMP 2011, and 30 (30) for LCO 2015
for modeling the full (partial) line profile.  The lower values of temperature needed
for the LCO 2015 dataset are due to the higher uncertainties of the spectral fluxes.
These large values of the temperature mean that the numerical noise
from specific placement of the point particles in position and velocity space
for a given set of parameter values is much less than the spectral flux errors
times $\sqrt{T}$.  
We check all results for convergence by comparing the inferred parameter distributions
from the first and second halves of the modeling run.

%%%%%%%%%%%%%%%%%%%%%%%%%%%%%%%%%%%%%%
%%%%%%%%%%%%%%%%%%%%%%%%%%%%%%%%%%%%%%
%%%%%%%%%%%%%%%%%%%%%%%%%%%%%%%%%%%%%%
%%%%%%%%%%%%%%%%%%%%%%%%%%%%%%%%%%%%%%

\section{Results}
 \label{sect_results}
We now present the results from applying our geometric and dynamical model for
the BLR to three velocity-resolved reverberation mapping datasets for Arp 151. 
As discussed in Section~\ref{sect_systematics}, one of the sources of systematic 
uncertainty in our analysis is the isolation of \Hb\ flux in the red wing, where
the choice of \feii\ template and possible contribution of \hei\ lines introduce
differences in the \Hb\ red wing that can exceed the spectral flux uncertainties.  
To address this, we present two limiting cases for all three datasets of modeling the
full \Hb\ red wing or approximately half of the red wing.  
After describing the results for the individual datasets, we then
combine the results to provide a joint inference on the BLR model parameters. 
The median and 68\% confidence intervals of the inferred BLR model
parameter values are given in Table~\ref{table_results}.

 \begin{deluxetable*}{lccccc}
%\tabletypesize{\scriptsize}
\tablecaption{Inferred BLR Model Parameter Values}
\tablewidth{0pt}
\tablehead{ 
\colhead{Parameter} & 
\colhead{LAMP 2008} &
\colhead{LAMP 2011} &
\colhead{LCO 2015} &
\colhead{Combined Datasets} &
\colhead{Standard Deviation} 
}
\startdata
$\tau_{\rm CCF}$ (days) &  $3.99^{+0.49}_{-0.68}$ &  $5.61^{+0.66}_{-0.84}$ &  $7.52^{+1.43}_{-1.06}$ &  $-$ &  $-$ \\ 
$r_{\rm max}$ (light days) &  $44.85$ &  $27.39$ &  $54.71$ &  $-$ &  $-$ \\ 
$r_{\rm mean}$ (light days) &   $4.07^{+0.42}_{-0.42}$ &   $6.77^{+0.52}_{-0.50}$ &   $7.09^{+1.42}_{-1.17}$ &  $-$ &  $-$ \\ 
$r_{\rm median}$ (light days) &   $2.74^{+0.65}_{-0.60}$ &   $5.35^{+0.50}_{-0.54}$ &   $4.32^{+1.16}_{-0.89}$ &  $-$ &  $-$ \\ 
$r_{\rm min}$ (light days) &   $0.57^{+0.20}_{-0.31}$ &   $0.71^{+0.40}_{-0.38}$ &   $0.50^{+0.52}_{-0.37}$ &   $0.65^{+0.15}_{-0.12}$ &  $0.08$ \\ 
$\sigma_{r}$ (light days) &   $3.89^{+0.74}_{-0.63}$ &   $5.62^{+1.03}_{-0.63}$ &   $8.03^{+2.02}_{-1.59}$ &  $-$ &  $-$ \\ 
$\tau_{\rm mean}$  (days)  &   $3.65^{+0.34}_{-0.38}$ &   $5.99^{+0.41}_{-0.35}$ &   $6.04^{+0.80}_{-0.86}$ &  $-$ &  $-$ \\ 
$\tau_{\rm median}$  (days)  &   $2.28^{+0.60}_{-0.53}$ &   $4.52^{+0.43}_{-0.36}$ &   $3.37^{+0.62}_{-0.55}$ &  $-$ &  $-$ \\ 
$\beta$ &   $1.14^{+0.26}_{-0.28}$ &   $0.90^{+0.14}_{-0.11}$ &   $1.21^{+0.15}_{-0.13}$ &   $1.01^{+0.17}_{-0.09}$ &  $0.13$ \\ 
$\theta_o$ (degrees) &   $24.6^{+ 5.5}_{- 7.8}$ &   $18.0^{+ 5.7}_{- 6.6}$ &   $22.2^{+ 9.4}_{-10.0}$ &   $21.8^{+ 2.7}_{- 5.4}$ &  $2.72$ \\ 
$\theta_i$ (degrees) &   $23.7^{+ 5.1}_{- 7.6}$ &   $15.1^{+ 3.7}_{- 5.0}$ &   $19.7^{+ 8.1}_{- 9.3}$ &   $15.3^{+ 3.9}_{- 5.2}$ &  $3.54$ \\ 
$\kappa$ &   $-0.23^{+0.28}_{-0.15}$ &   $-0.06^{+0.32}_{-0.30}$ &   $0.10^{+0.28}_{-0.31}$ &   $-0.19^{+0.29}_{-0.15}$ &  $0.13$ \\ 
$\gamma$ &   $3.60^{+1.01}_{-1.38}$ &   $2.80^{+1.39}_{-1.13}$ &   $3.78^{+0.89}_{-1.19}$ &   $3.81^{+0.93}_{-0.97}$ &  $0.42$ \\ 
$\xi$ &   $0.22^{+0.33}_{-0.16}$ &   $0.23^{+0.18}_{-0.13}$ &   $0.28^{+0.23}_{-0.17}$ &   $0.17^{+0.16}_{-0.09}$ &  $0.03$ \\ 
$\log_{10}(M_{\rm BH}/M_\odot)$ &   $6.66^{+0.26}_{-0.17}$ &   $6.93^{+0.33}_{-0.16}$ &   $6.92^{+0.50}_{-0.23}$ &   $6.82^{+0.09}_{-0.09}$ &  $0.13$ \\ 
$f_{\rm ellip}$ &   $0.21^{+0.26}_{-0.15}$ &   $0.30^{+0.13}_{-0.17}$ &   $0.18^{+0.16}_{-0.13}$ &   $0.18^{+0.14}_{-0.13}$ &  $0.05$ \\ 
$f_{\rm flow}$ &   $0.25^{+0.17}_{-0.18}$ &   $0.25^{+0.16}_{-0.18}$ &   $0.26^{+0.16}_{-0.17}$ &   $0.27^{+0.15}_{-0.20}$ &  $0.004$ \\ 
$\theta_e$ (degrees) &   $14.8^{+16.9}_{-10.2}$ &   $13.6^{+16.0}_{- 9.1}$ &   $20.1^{+18.1}_{-13.5}$ &   $10.1^{+ 8.8}_{- 6.2}$ &  $2.83$ \\ 
$\sigma_{\rm turb}$ &   $0.012^{+0.041}_{-0.010}$ &   $0.046^{+0.034}_{-0.040}$ &   $0.010^{+0.029}_{-0.008}$ &   $0.044^{+0.030}_{-0.037}$ &  $0.017$ \\ 
\enddata
\tablecomments{Median values and 68\% confidence intervals for the main BLR geometry and dynamics model parameters.
The values for the column Combined Datasets are measured from the joint posterior PDFs for the three datasets.  The values
for the last column, Standard Deviation, are the standard deviation of the values for the three individual datasets.  The fixed 
values for $r_{\rm max}$ used in BLR modeling are also shown as well as the centroid time lag $\tau_{\rm CCF}$ from cross-correlation function analysis from \citet{bentz09}, Barth et al. (in preparation), and \citet{valenti15} for LAMP 2008, LAMP 2011, and LCO 2015, respectively.  All values are redshift-corrected to the AGN rest-frame.
\label{table_results}}
\end{deluxetable*}

\begin{figure*}[h!]
\begin{center}
\includegraphics[scale=0.5]{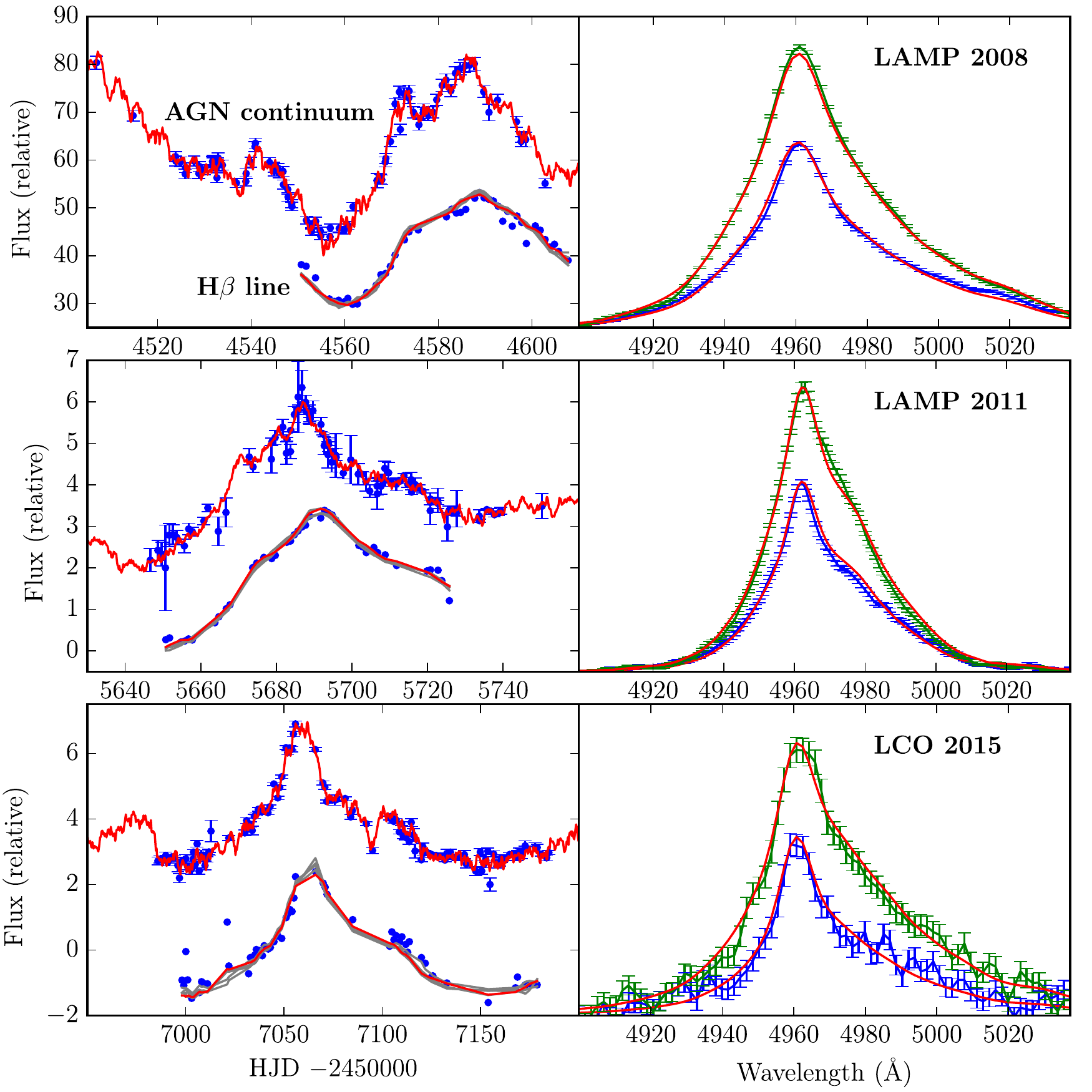}  % left, bottom, right, top
\includegraphics[scale=0.5]{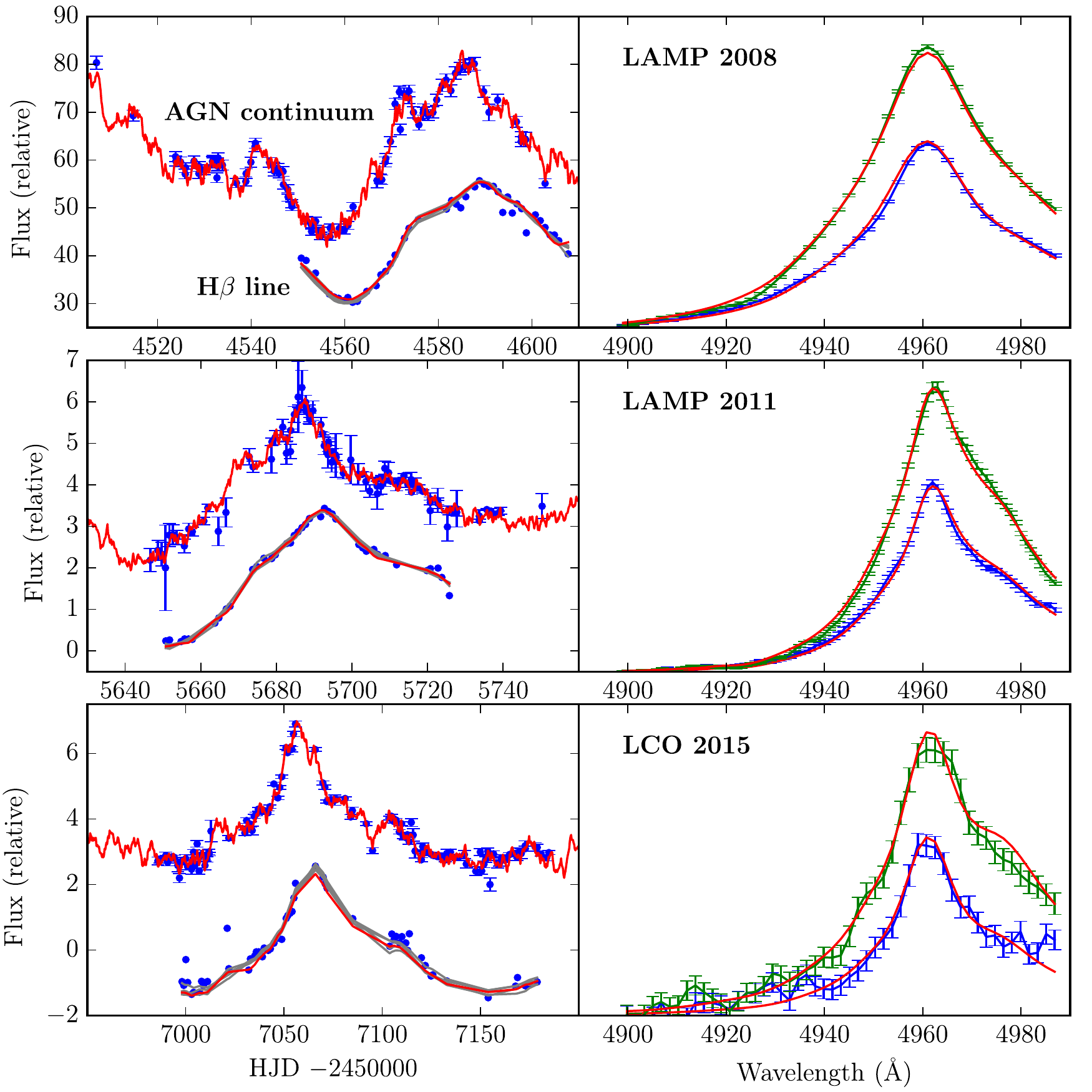}  % left, bottom, right, top
\caption{Model fits to the emission line profile for the LAMP 2008 (top panels), 
LAMP 2011 (middle panels), and LCO 2015 (bottom panels) datasets.  
The set of six lefthand-side panels show the model fits to the full emission line
profile and the set of six righthand-side panels show the model fit to the partial
emission line profile when part of the red wing is excluded.
On the left, for each dataset and for both the full and partial fits to the data, 
we show the AGN continuum light curve at the top (blue points and error-bars) with
an example of a Gaussian process continuum light curve model (red line)
and the integrated \Hb\ emission line light curve at the bottom (blue points) with examples
of the model fit drawn from the posterior PDF (red and grey lines).
On the right we show two examples of the model fit (red lines) to individual emission
line profiles (blue and green error-bars and lines).
\label{fig_allfit}}
\end{center}
\end{figure*}

\subsection{Inference for Individual Datasets}

\begin{figure*}[h!]
\begin{center}
\includegraphics[trim={3cm 0 3cm 0},clip, scale=0.55]{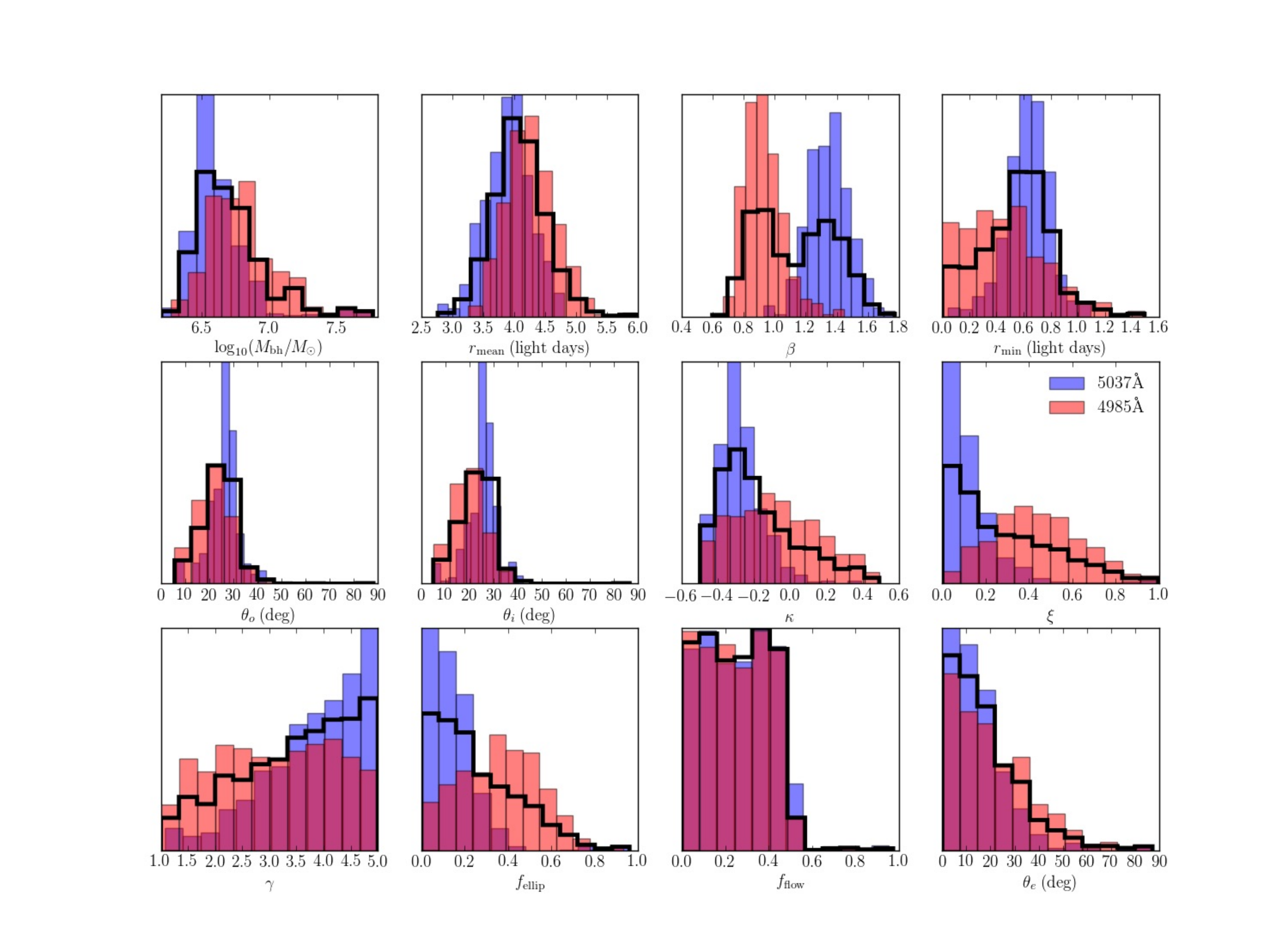}  % left, bottom, right, top
\caption{Inferred posterior PDFs of the main BLR modeling parameters for the LAMP 2008 dataset from
modeling the full emission line profile out to 5037~\AA\ (blue histograms) and from excluding part of the red wing
after 4987~\AA\ (red histogram).  An equal mixture of the red and orange posterior PDFs is given by the solid black 
line histogram.
\label{fig_lamp08}}
\end{center}
\end{figure*}

\begin{figure*}[h!]
\begin{center}
\includegraphics[trim={3cm 0 3cm 0},clip, scale=0.55]{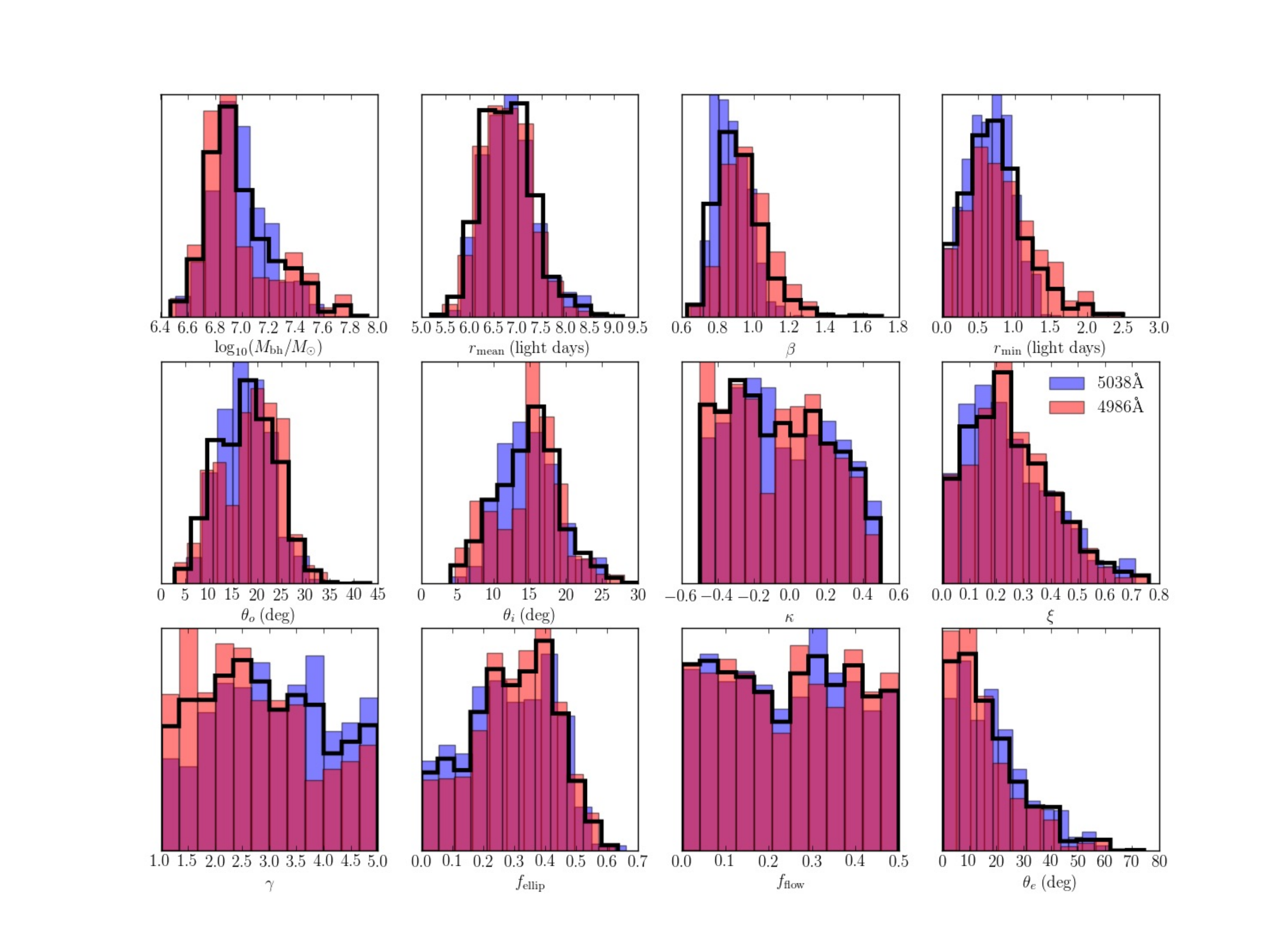}  % left, bottom, right, top
\caption{Same as Figure~\ref{fig_lamp08} for the LAMP 2011 dataset. \label{fig_lamp11}}
\end{center}
\end{figure*}

\begin{figure*}[h!]
\begin{center}
\includegraphics[trim={3cm 0 3cm 0},clip, scale=0.55]{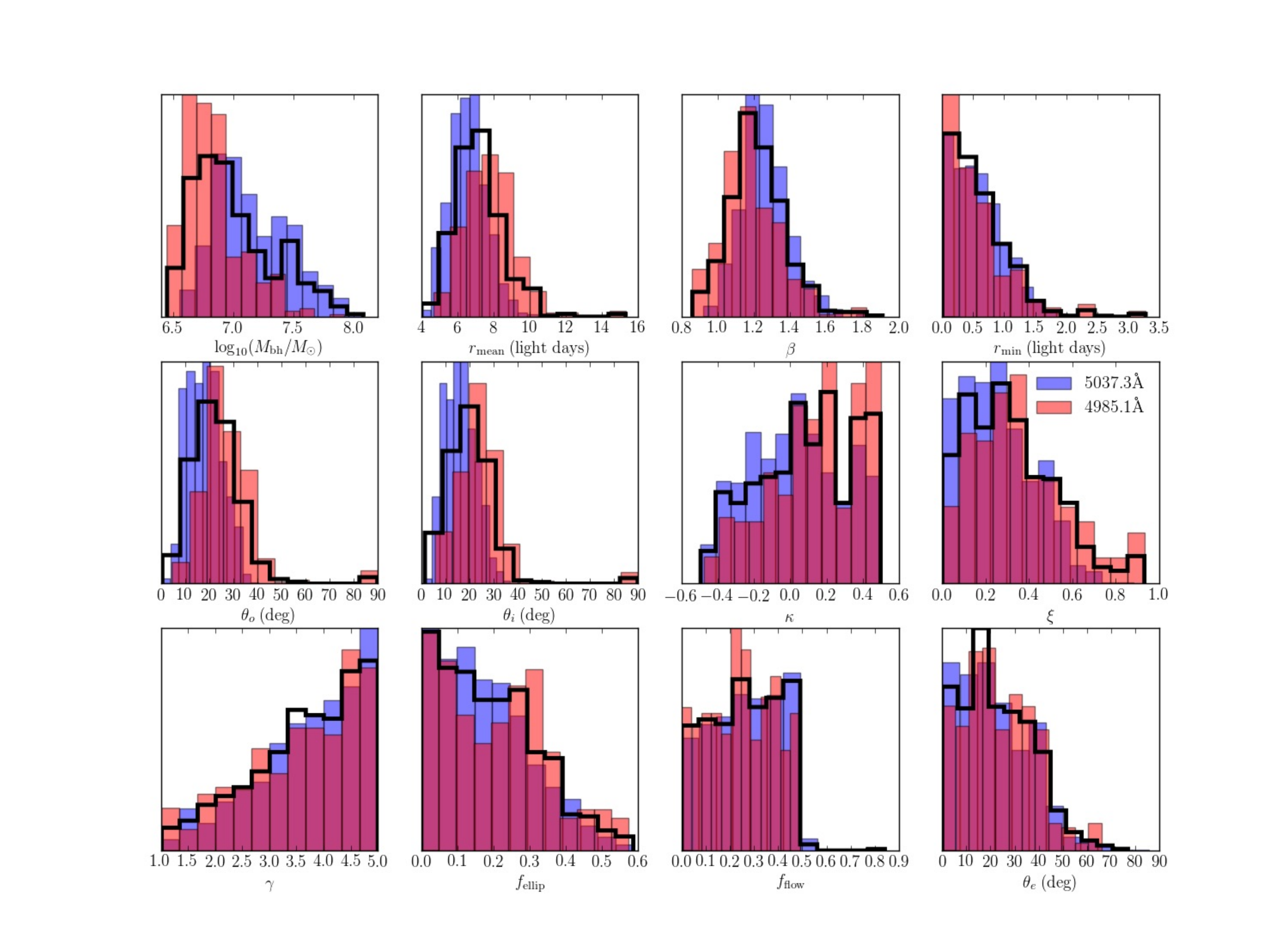}  % left, bottom, right, top
\caption{Same as Figure~\ref{fig_lamp08} for the LCO AGN Key Project 2015 dataset. \label{fig_agnkey}}
\end{center}
\end{figure*}

The LAMP 2008 dataset for Arp 151 has previously been analyzed using the 
BLR modeling approach by \citet{brewer11} and \citet{pancoast14b}.  
Our analysis differs from that of \citet{pancoast14b} in three ways.  
First, we use an updated model for the BLR that includes the 
systematic central wavelength of the broad emission line 
as a free parameter and a maximum outer radius for the BLR.
Second, the \Hb\ emission line profile has been isolated from the spectrum
using a different spectral decomposition code and the \feii\ template from
\citet{kovacevic10} instead of from \citet{boroson92}.
Third, in addition to modeling the \Hb\ emission line profile over the wavelength range of 
$4899 - 5037$~\AA\ in the observed frame, we also exclude approximately half of the red wing and model the
\Hb\ emission line profile over the range of $4899 - 4985$~\AA.
For both wavelength ranges of the data, the BLR model is able to capture the shape of the \Hb\ emission
line profile and the large-scale changes of the integrated \Hb\ line flux, as shown in the top panels
of Figure~\ref{fig_allfit}.  

The LAMP 2011 dataset for Arp 151 is the second 
in the sample to be analyzed using the BLR modeling approach, after Mrk 50 \citep{pancoast12}.  
Compared to LAMP 2008, the LAMP 2011 dataset showed similarly high variability, but with increased
instrumental resolution.  Again, we model both the full \Hb\ emission line profile over the range
of $4899 - 5038$~\AA\ and also model a truncated version of the \Hb\ emission line profile excluding approximately half of the red wing, 
for a wavelength range of $4899 - 4986$~\AA.  These wavelength ranges are very close
to those used for the LAMP 2008 dataset, although with a different pixel scale.
As for LAMP 2008, the BLR model is able to fit both the shape of the \Hb\ emission line profile
and follow the changes in the integrated \Hb\ line flux as a function of time.
Examples of the model fit to the data are shown in the middle panels of Figure~\ref{fig_allfit}. 

The dataset for Arp 151 from 2015 is the first 
velocity-resolved reverberation mapping result from the LCO AGN Key Project \citep{valenti15}.  
Compared to the LAMP
datasets for Arp 151, the LCO 2015 dataset has similarly high levels of variability and instrumental
resolution between the two LAMP datasets, but significantly lower signal-to-noise (S/N) ratio for the spectroscopy.
We model both the full \Hb\ emission line profile between $4899.8 - 5037.3$~\AA\ as well as excluding approximately half of the red wing to model the line profile between $4899.8 - 4985.1$~\AA. 
Other than two highly discrepant data points in the \Hb\ light curve, the BLR model is able to fit the \Hb\ emission
line profile shape as well as match the overall variability of the integrated \Hb\ line flux, as shown by
the bottom panels in Figure~\ref{fig_allfit}. 

We show the inferred posterior PDFs for some of the key BLR model parameters in Figures~\ref{fig_lamp08}, 
\ref{fig_lamp11}, and \ref{fig_agnkey} for the LAMP 2008, LAMP 2011, and LCO 2015 datasets respectively.
To ease comparison between the results for modeling the two different wavelength
ranges, we show the posterior PDFs from modeling the emission line profile out to $\sim 5037$~\AA\ in blue
and the posterior PDFs from modeling the emission line profile out to only $\sim 4985$~\AA\ in orange
 (areas where the two posterior PDFs overlap appears red).  
An equal 50/50 mixture of the orange and blue posterior PDFs is shown by the thick black histogram
and used to measure the median and 68\% confidence intervals for the 
BLR model parameters listed in the first three columns of Table~\ref{table_results} and described below.

For the geometry of the \Hb-emitting BLR we infer a thick disk viewed close to face-on 
with a disk opening angle of $\theta_o=$\,\lampaThetao\ (\lampbThetao, \lcoThetao) deg and an inclination angle with
respect to the observer's line of sight of $\theta_i=$\,\lampaThetai\ (\lampbThetai, \lcoThetai) deg ($0=$ face-on), 
for LAMP 2008 (LAMP 2011, LCO 2015).  
The distribution of emission decreases exponentially or steeper as a function of radius, with $\beta =$\,\lampaBeta\ (\lampbBeta, \lcoBeta) for 
LAMP 2008 (LAMP 2011, LCO 2015), 
and the median radius of emission changes by almost a factor of two, from $r_{\rm median}=$\,\lampaRmedian\ light days for LAMP 2008 to $r_{\rm median}=$\,\lampbRmedian\ (\lcoRmedian) light days for LAMP 2011 (LCO 2015).
The asymmetry of emission is inferred to varying degrees by the three datasets.
The LAMP 2008 data prefer BLR models where the \Hb\ emission comes more 
from the far side of the BLR with respect to the observer
($\kappa=$\,\lampaKappa), although more emission from the near side of the 
BLR is not ruled out and the LAMP 2011 and LCO 2015
datasets have no strong preference.  
All three datasets prefer BLR models where the disk mid-plane is partially or completely opaque, 
with $\xi=$\,\lampaXi\ (\lampbXi, \lcoXi) for LAMP 2008 (LAMP 2011, LCO 2015), 
and a completely transparent disk mid-plane is ruled out for LAMP 2011.
Finally, the LAMP 2008 and LCO 2015 data have a slight preference for BLR models with more emission from the faces of
the disk ($\gamma=$\,\lampaGamma, \lcoGamma, respectively).

For the dynamics of the \Hb-emitting BLR we infer a combination of near-circular
elliptical and inflowing orbits for the emitting gas, 
with the fraction of near-circular elliptical orbits given by $f_{\rm ellip}=$\,\lampaFellip\ (\lampbFellip, \lcoFellip) 
for LAMP 2008 (LAMP 2011, LCO 2015).  
The remaining gas is inflowing with $f_{\rm flow}=$\,\lampaFflow\ (\lampbFflow, \lcoFflow), where $0 \le f_{\rm flow} \le 0.5$ indicates inflow, and the gas is anywhere from half gravitationally bound on radial orbits ($\theta_e \to 0$ deg) 
to mostly gravitationally bound on 
both tangential and radial orbits (out to $\theta_e \sim 50$ deg) with $\theta_e=$\,\lampaThetae\ (\lampbThetae, \lcoThetae) for LAMP 2008 (LAMP 2011, LCO 2015).
There is also a negligible contribution from macroturbulent velocities with
$\sigma_{\rm turb}=$\,\lampaSigmaturb\ (\lampbSigmaturb, \lcoSigmaturb).

Finally, the BH mass setting the scale of the velocity 
field is inferred to be $\log_{10}($\mbh$)=$\,\lampaMbh\ (\lampbMbh, \lcoMbh).
While there is a difference of almost 0.3 dex between the BH mass measured for LAMP 2008 
compared to the other two, this difference is within the statistical measurement uncertainties and the posterior PDFs
significantly overlap as seen in Figure~\ref{fig_final}.  
 The standard deviation in the three BH mass measurements is significantly smaller, at only 0.13 dex.
It is also interesting to compare these BH mass values from BLR modeling with values of the virial product (VP) 
calculated in traditional reverberation mapping analysis, 
as shown in the top panel of Figure~\ref{fig_evolution}.
The VP is related to the BH mass by \mbh$=f\times$VP and 
$\log_{10}($VP$)= 6.086^{+0.075}_{-0.091}$, $5.778^{+0.061}_{-0.071}$, and $6.053^{+0.077}_{-0.094}$ for the LAMP 2008 \citep{bentz09}, LAMP 2011 (Barth et al., in preparation), and LCO 2015 \citep{valenti15} datasets, respectively.
The standard deviation in the VP values is 0.14 dex, very similar to the value for BLR modeling.  
While the statistical uncertainties from calculating the VP are quite small, generally $<0.1$~dex,
calculating a VP BH mass requires the additional assumption of choosing a value for the virial factor $f$.  
Traditionally, an average value of $f$ is derived by matching the reverberation mapping sample to the
quiescent galaxy sample \msigma\ relation \citep{onken04, collin06, woo10, greene10b, graham11, park12b, woo13, grier13b, woo15, batiste17}.
However, the scatter in the \msigma\ relation is measured to be at least 0.4~dex \citep{park12b}, suggesting that the total uncertainty of VP BH masses
could be as large as $\sim 0.4$ dex for individual AGN \citep[as discussed by][]{peterson14}.
This means that the additional uncertainty from the unknown value of $f$ for individual AGN is significant and must be
included when comparing BH mass measurement techniques. 
These results show that BLR modeling for Arp 151 generally provides greater precision than VP BH masses with uncertainties $<0.4$~dex, as well as
constraints on BLR structure, and thus $f$, independent of the \msigma\ relation.

The inferred structure of the \Hb-emitting BLR in Arp 151 for the
datasets described above combines the results for both
modeling the full emission line profile and excluding half of the red wing.
However, some of the BLR model parameters are sensitive to how much of the red wing is modeled.
As evident from Figure~\ref{fig_lamp08}, 
the parameters in best agreement for LAMP 2008 include
$r_{\rm mean}$, $\theta_o$, $\theta_i$, $f_{\rm flow}$, $\theta_e$, and
\mbh, while $\beta$, $r_{\rm min}$, $\kappa$, $\xi$, $\gamma$,
and $f_{\rm ellip}$ are more dissimilar.  Excluding half of
the red wing provides weaker constraints on the emission asymmetry parameters
$\kappa$, $\xi$, and $\gamma$, but also on the dynamics through $f_{\rm ellip}$ and 
larger inferred uncertainties for
\mbh.  Since \mbh\ is strongly correlated with $\theta_o$ and $\theta_i$, they 
also have larger inferred uncertainties when excluding half the red wing.
Overall, modeling the full red wing is significantly more constraining for LAMP 2008, but 
the results are generally consistent with one another, with the largest discrepancy
in the inferred values of $\beta$, for which the posterior PDFs only slightly overlap.
On the other hand, modeling the full
red wing does not provide significantly better constraints on BLR structure for the LAMP 2011 dataset, since the posterior PDFs
from modeling the full and partial red wing
almost completely overlap one another (Figure~\ref{fig_lamp11}). 
Similarly, modeling the full
red wing for the LCO 2015 dataset does not provide significantly better constraints (Figure~\ref{fig_agnkey}).  
However, there is a slight offset in the
posterior PDFs for  \mbh, $r_{\rm mean}$, $\beta$, $\theta_o$, $\theta_i$, and $\xi$,
such that modeling the full red wing leads to larger values of the BH mass
and, through the tight correlations with the inclination and opening angles, a 
thinner and more face-on disk.

\subsection{Inference from Combining the Three Datasets  \label{sect_results_all}}

\begin{figure*}[h!]
\begin{center}
\includegraphics[trim={3cm 0 3cm 0},clip, scale=0.55]{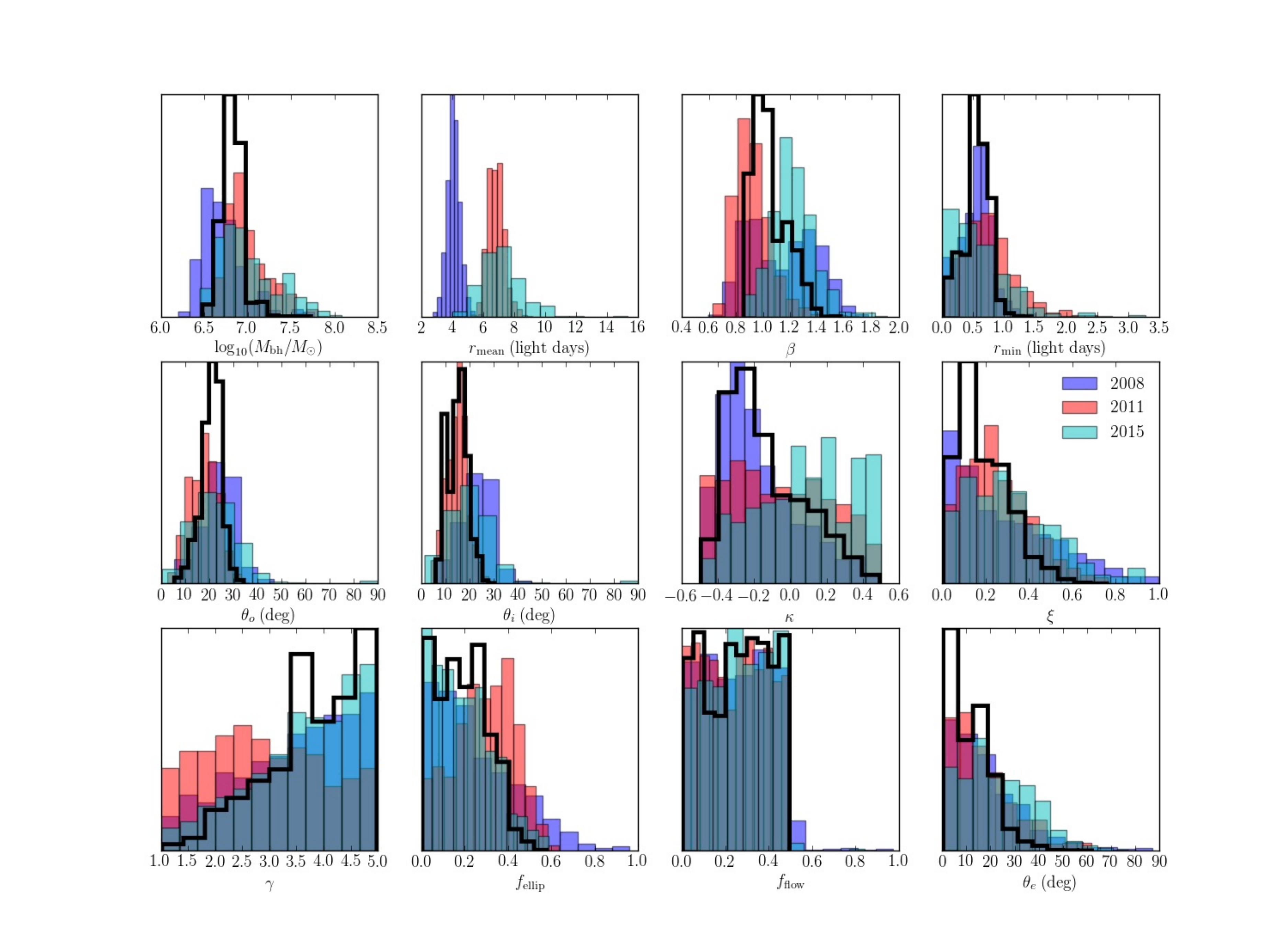}  % left, bottom, right, top
\caption{Inferred posterior PDFs of the main BLR modeling parameters for the LAMP 2008 (blue histogram), LAMP 2011
(red histogram), and LCO 2015 (cyan histogram) datasets.  The posterior PDFs for each dataset consist
of an equal mixture of posterior samples from modeling the full emission line profile and from modeling a partial 
emission line profile that excludes part of the red wing.  
We also show the joint inference (black line histogram) from
multiplying the inferred likelihood distributions for the three datasets together.
\label{fig_final}}
\end{center}
\end{figure*}

We now combine the results for the three datasets for Arp 151 to obtain a joint 
inference on the BLR model parameters.  Some parameters, such
as the BH mass, we expect to stay constant.  
Other parameters, such as the mean radius or other
measurements of the radial size of the BLR, we expect to change in response to 
variations in the AGN luminosity.  The evolution of BLR structure over the seven year time
period spanned by the three datasets is discussed in Section~\ref{sect_evolution}.

We start by defining the set of posterior samples for each dataset individually.  As
described in the previous section, we marginalize over the choice to model the full or partial \Hb\ red wing by making an equal 50/50
mixture of posterior samples from the results for each case.  Since the posterior PDFs from modeling the full or partial \Hb\ red wing never perfectly overlap, this addition of posterior PDFs has a general broadening effect on the final posteriors for each dataset.  
The equal mixtures of posterior samples for each dataset (LAMP 2008, LAMP 2011, LCO 2015) are shown by
the solid black histograms in Figures~\ref{fig_lamp08}, \ref{fig_lamp11}, and \ref{fig_agnkey}
and shown over-plotted in blue, red, and cyan, respectively, in Figure~\ref{fig_final}.
The next step is to combine the independent constraints provided by each of the three datasets to create a joint inference
on the BLR model parameters.
This is done by multiplying together the inferred likelihood functions for
each dataset before applying the prior probabilities to produce the joint inference posterior PDF.  
In order to multiply likelihood functions made of discrete samples, the posterior samples for each dataset are first divided by their prior probability function for priors that are not flat in the parameter (e.g. for parameters such as BH mass that have priors that are flat in the log of the parameter).
Then the likelihood samples are placed in 100 bins and the binned likelihood functions are multiplied before applying the prior probability functions again.  
The joint inference posterior PDFs are shown by
the solid black histograms in Figure~\ref{fig_final} and are used to calculate the median and 68\% confidence
intervals given in the fourth column of Table~\ref{table_results}.
Unlike the addition of posterior PDFs, which generally widens the distributions and increases
the uncertainties in the inferred parameter values, 
multiplying the likelihood functions for the three independent datasets shrinks the final 
posterior PDFs and decreases the uncertainties on the inferred parameter values.  

Overall, constraints on the \Hb-emitting BLR geometry are consistent between the three datasets,
with the exception of the size of the BLR, which grows by a factor of almost two between the 
LAMP 2008 and LAMP 2011 datasets.  Due to almost no overlap between the posterior PDFs of
the mean and median radius and time lag between LAMP 2008 and both LAMP 2011 and LCO 2015,
we do not calculate the combined posteriors for Figure~\ref{fig_final} or Table~\ref{table_results}.
For the asymmetry of the close to face-on, thick disk ($\theta_o=$\,\totThetao\ deg and $\theta_i=$\,\totThetai\ deg), 
we infer a preference for a mostly opaque 
disk mid-plane ($\xi=$\,\totXi) with a completely transparent mid-plane ruled out, 
more emission from the faces of the disk ($\gamma=$\,\totGamma), and
a slight preference for more emission back towards the central ionizing source ($\kappa=$\,\totKappa).

The constraints on the \Hb-emitting BLR dynamics are also consistent between the three
datasets, with less than 50\% of the emitting material in near-circular elliptical orbits 
($f_{\rm ellip}=$\,\totFellip) and the
remaining gas in mostly radial inflowing orbits ($f_{\rm flow}=$\,\totFflow\ and $\theta_e=$\,\totThetae\ deg).
Finally, the BH mass has a combined inference of $\log_{10}($\mbh$)=$\,\totMbh, more precise than
any of the individual black hole mass measurements for Arp 151 for a single dataset
by $\sim 0.1$ dex and the most precise BH mass from BLR modeling to date.

%%%%%%%%%%%%%%%%%%%%%%%%%%%%%%%%%%%%%%
%%%%%%%%%%%%%%%%%%%%%%%%%%%%%%%%%%%%%%
%%%%%%%%%%%%%%%%%%%%%%%%%%%%%%%%%%%%%%
%%%%%%%%%%%%%%%%%%%%%%%%%%%%%%%%%%%%%%

\section{Discussion} \label{sect_discussion}

\subsection{Systematic Uncertainties in the BLR Modeling Approach  \label{sect_systematics}}
Previous BLR modeling analysis \citep{brewer11, pancoast12, pancoast14b} has 
focused on providing statistical uncertainties in the inferred structure of the BLR as provided by
either MCMC or diffusive nested sampling algorithms.  Diffusive nested sampling, in
particular, provides robust statistical uncertainties for the BLR model parameters even 
in the case of tight parameter degeneracies and multimodal posterior PDFs.  
However, there are additional uncertainties not captured by sampling statistics, 
including what we define as \Hb\ line emission and what model we use for the BLR.
We will discuss these sources of systematic uncertainty in the following sections.

\subsubsection{Spectral Decomposition}
One of the most important assumptions we make in the process of modeling the BLR is that we can robustly isolate the broad line flux from the full spectrum using spectral decomposition.  However, this requires the choice of both individual templates for certain spectral components as well as which emission lines or components are present in the data.  The two most difficult choices to make for the three Arp 151 datasets are 1) which \feii\ template to use and 2) whether \hei\ at 4471, 4922, and 5016~\AA\ rest wavelength is noticeably present in the spectrum.  Both of these spectral components overlap the red wing of \Hb\ and can be difficult to disentangle.  To quantify the uncertainty introduced by our choice of \feii\ template and exclusion of \hei\, we compared the standard deviation of spectral decomposition solutions for all three \feii\ templates with and without \hei, $\Delta f_{\rm decomp}$, to the spectral flux errors, $\sigma_{\rm flux}$.  
The ratio of $\Delta f_{\rm decomp}/\sigma_{\rm flux}$ is 4 (5, 2) times larger for the \Hb\ red wing between $\sim 4985-5037$~\AA\ compared to the rest of the line and the median value of this ratio for the red wing is 2.4 (2.1, 0.4) for LAMP 2008 (LAMP 2011, LCO 2015).  
This shows that, at least for the LAMP 2008 and 2011 datasets and for some epochs of LCO 2015, the choices of \feii\ template and presence of \hei\ in the spectral decomposition do meaningfully affect the \Hb\ line profile, leading to a source of systematic uncertainty in the isolation of \Hb\ flux specifically in the red wing.  
Looking forward, we can turn this systematic uncertainty into an additional statistical uncertainty in future BLR modeling work by inferring a posterior sample of spectral decompositions instead of a single best-fit dataset, thereby marginalizing over the choice of \feii\ template and presence of \hei\ in the data.

\subsubsection{The Full or Partial Emission Line Profile}
One way to probe the magnitude of the systematic uncertainty introduced from spectral decomposition is to compare the BLR properties inferred for the full line profile to the properties inferred when the \Hb\ red wing is excluded, as we describe in Section~\ref{sect_results}.  This comparison shows that most, but not all, BLR model parameters are consistently inferred.  However, this comparison requires excluding just over a third of the line profile, probing another source of systematic uncertainty: how much of the line profile we model.  In addition to excluding the \Hb\ red wing, we also tried excluding either the \Hb\ blue wing or the center of the emission line profile.   
Masking each region results in slightly different inferences on BLR model parameters that are generally consistent with one another, with the two cases of modeling the full line and excluding the red wing providing a typical level of difference.  However, given the good agreement in spectral decomposition in the blue wing and center of the emission line profile, we do not include posterior samples from these runs in our final inference for each dataset, as this would generally widen the inferred posteriors due to modeling a smaller fraction of the data.  Excluding the red wing (blue wing, center) of \Hb\ results in larger uncertainties for inferred BLR model parameters by $10\pm20$\% ($20\pm40$\%, $50\pm130$\%) for LCO 2015.  For LAMP 2008 and 2011, excluding the red wing (center) of \Hb\ results in larger uncertainties by $40\pm70$\% ($40\pm50$\%) and $3\pm30$\% ($20\pm50$\%), respectively. This has an important implication for future velocity-resolved reverberation mapping programs focused on wide \Hb\ emission lines where the line wings may be heavily contaminated by other spectral features (e.g. variable \heii\ in the blue wing and \hei\ and variable \feii\ in the red wing): to obtain the smallest statistical uncertainties from BLR modeling, robust identification of \Hb\ flux is crucial all across the line profile so that no portion of the line profile needs to be excluded.  

\subsubsection{Using a Simple Model \label{sect_simple_model}}
There are many facets of our simply parameterized phenomenological model for the BLR that could introduce systematic uncertainties in the inferred model parameters.  
Here we discuss possible sources of systematic uncertainty from 1) recent changes to the BLR model, 2) correlations between model parameters, and 3) basic assumptions about the BLR physics.  

There are two main changes to the BLR model that were made in the process of analyzing the three datasets for Arp 151.  First, the central wavelength of \Hb\ emission is now a free parameter with a Gaussian prior of standard deviation 1~\AA\ (LAMP 2008) or 4~\AA\ (LAMP 2011, LCO 2015).  While using a narrower prior for the central wavelength sometimes leads to more precise inferences on some BLR model parameters, the effects are generally small: 
comparing results from using Gaussian priors of width 1~\AA\ or 4~\AA\ makes 
differences between the inferred BLR model parameters that are $<40$\% ($<40$\%, $<20$\%) of the inferred statistical uncertainties for LAMP 2008 (LAMP 2011, LCO 2015), with typical values $<20$\% for all three datasets.  
The second change to the BLR model is that there is now a set maximum outer radius of BLR emission, with values given in Table~\ref{table_results} for each dataset.  To test whether the choice of maximum outer radius affects the inferred BLR model parameters, we analyzed each dataset using a value for the maximum radius that was approximately twice as large.  This test showed that the choice of maximum radius has a larger effect than the choice of Gaussian prior width for the central \Hb\ wavelength, but the differences between the inferred BLR model parameters are still less than $<60$\% ($<50$\%, $<20$\%) of the inferred statistical uncertainties for LAMP 2008 (LAMP 2011, LCO 2015).  While adding the maximum radius as a free parameter would be the best way to incorporate this source of systematic uncertainty into the inferred statistical uncertainty, it comes with a computational cost of requiring a longer extrapolated AGN continuum light curve.

Another source of systematic uncertainty comes from correlations between model parameters, as described in detail by \citet{grier17}.
On one hand, the ability of our flexible BLR model to use multiple, discrete combinations of parameter values to generate the same distribution of point particles in position and velocity space inflates the statistical uncertainties on the inferred model parameters compared to the true uncertainty in specific point particle distributions.  One example of this is the degeneracy between solutions with $f_{\rm ellip} \to 1$ and solutions with any value of $f_{\rm ellip}$ or $f_{\rm flow}$ when $\theta_e \to 90$ deg; in both cases the dynamics are dominated by near-circular elliptical orbits.  
On the other hand, however, degeneracies between model parameters can illuminate systematic uncertainties on a larger scale through the identification of parameter correlations that are unphysical for a sample of AGN.  The best example identified so far is a correlation between the inclination and opening angles for the LAMP 2008 \citep{pancoast14b} and AGN10 \citep{grier17} samples.  Tests with simulated data confirm that in order to produce single-peaked emission line profiles as observed for LAMP 2008 and AGN10, our model for the BLR requires that $\theta_o \gtrsim \theta_i$.  This places an effective prior on $\theta_o$ between $\theta_i$ and 90 deg.  
The fact that BLR modeling 
infers values of $\theta_o \sim \theta_i$ suggests that in reality $\theta_o \lesssim \theta_i$.  
While it is difficult to quantify how much the values of $\theta_o \sim \theta_i$ are pulled to the true values of either $\theta_o$ or $\theta_i$, tests with simulated data suggest that the shape of the velocity-resolved transfer function is more sensitive to values of $\theta_i$.  To obtain fully independent inferences on the inclination and opening angles we will need to include methods for creating a single-peaked line profile for values of $\theta_o < \theta_i$ in future BLR models. 

Finally, the BLR model makes many assumptions about the physics of gas in the inner regions of AGN that could add significant systematic uncertainty to our results.  While many of these assumptions are discussed in more detail by \citet{pancoast14a}, recent analysis of a multiwavelength reverberation mapping campaign of NGC 5548 has brought into question the validity of our assumption that the AGN continuum can be treated as coming from a point source \citep{edelson15, fausnaugh16}.  Specifically, the dataset for NGC 5548 provides a robustly measured time lag between the UV continuum at 1367~\AA\ as measured using the {\it Hubble Space Telescope} and the $\it V$-band optical continuum of $2.04^{+0.22}_{-0.20}$ days \citep{fausnaugh16}, which is $\sim 50$\% of the time lag between the 5100~\AA\ optical continuum and \Hb\ line emission \citep{pei17}.  At first glance this result suggests that it is problematic both to assume the continuum is coming from a point source and that the optical light curve is a reasonable proxy for the true ionizing continuum in the UV, leading to BH masses from BLR modeling that are too small by a factor of $\tau_{\rm opt}/\tau_{\rm UV}$, where $\tau_{\rm opt}$ ($\tau_{\rm UV}$) is the time lag between the optical (UV) continuum and \Hb.  However, it should be noted that NGC 5548 deviated from the usual $r_{\rm BLR}-L_{\rm AGN}$ relation
during this campaign with an \Hb\ BLR size smaller by a factor of $\sim 4$ compared to what the AGN luminosity predicted \citep{pei17}.  If NGC 5548 had not deviated from the $r_{\rm BLR}-L_{\rm AGN}$ relation then the time lag between the UV and optical continuum would only be $\sim 10$\% of the time lag between the optical continuum and \Hb, on the order of the statistical uncertainties inferred for the mean radius or time lag from BLR modeling.  

Given the change in BLR size between the three datasets for Arp 151 and the fact that we infer a larger BH mass for the datasets with larger BLR radii, we can estimate what time lag between the AGN continuum UV- and optical-emitting accretion disk regions would be required to explain the difference in inferred BH mass.  Using the median BH mass values listed in Table~\ref{table_results} suggests that the UV-optical time lag would need to be 0.85 times the optical-\Hb\ time lag if the larger BH mass we infer is correct.  If this were the case, however, even the datasets with larger inferred BLR radii would still be significantly affected by the UV-optical time lag.  If we instead solve for the UV-optical time lag with unknown true BH mass, we estimate UV-optical lags of $30-50$ days depending on what measurement of the radius or lag is used, resulting in a true BH mass of $10^{7.84-7.95}$\msun, an order of magnitude greater than what we infer.  Given the large difference between the estimated UV-optical lag and the inferred optical-\Hb\ lag, it is unlikely that the difference in inferred BH mass is primarily due to a violation of our assumption about an AGN continuum point source.
Additional simultaneous UV and optical reverberation mapping campaigns focused on other AGN will be necessary to determine how widely the results for NGC 5548 can be applied to the larger reverberation mapping sample.

\subsection{Evolution of the Broad Line Region  \label{sect_evolution}}

\begin{figure}[h!]
\begin{center}
\includegraphics[scale=0.6]{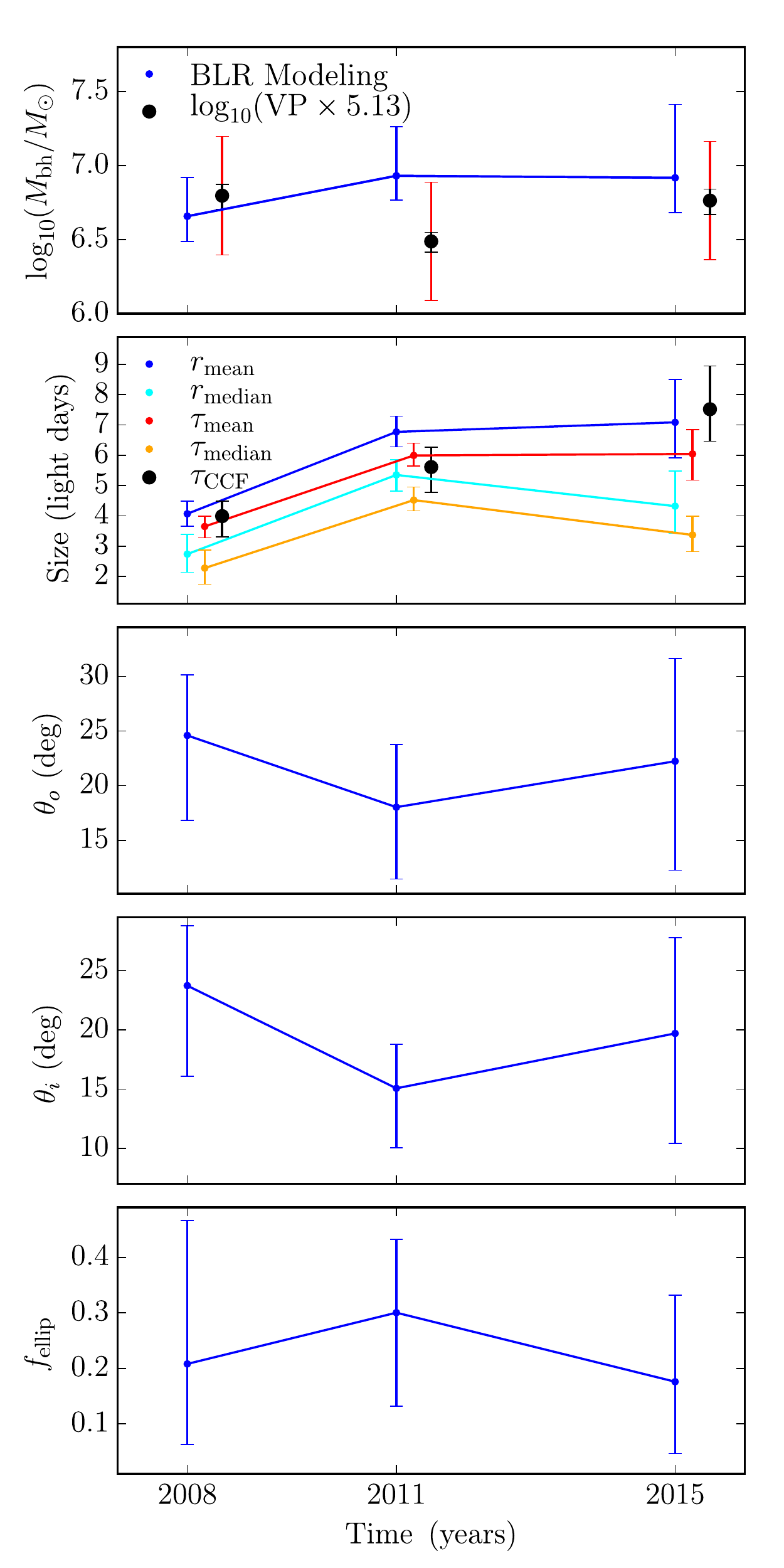}  % left, bottom, right, top
\caption{Evolution of key BLR model parameters over the time spanned by the three Arp 151 datasets.  
The BH mass from BLR modeling is shown by the blue points in the first panel.  For comparison, the virial product (VP) masses are also shown by black circles, where the small black error-bars are from the statistical uncertainties from measuring the VP and the larger red error-bars of 0.4 dex represent an estimate of the systematic uncertainties from using a mean value of $f=5.13$ measured by \citet{park12b}.  
Four model parameters describing the BLR size are shown in the second panel, including the mean and median radius (in blue and cyan, respectively) and the mean and median time lag (in red and orange, respectively).  The BLR size as measured by the model-independent CCF time lag is also shown by the black points for comparison.
 \label{fig_evolution}}
\end{center}
\end{figure}

The three datasets for Arp 151 span a time period of seven years, which is the orbital time for BLR gas at radii of $\sim 4$ light days from the BH.  We now discuss any evidence of evolution in BLR structure over this time, with the time dependence of key BLR model parameters shown in Figure~\ref{fig_evolution}.   As presented in Section~\ref{sect_results_all}, the largest difference between inferred BLR parameters for the three datasets is the inference in the radial size of the BLR, such that the mean and median radii and time lags for LAMP 2011 and LCO 2015 are a factor of almost $2$ greater than for LAMP 2008.  
As illustrated in the second panel of Figure~\ref{fig_evolution}, the change in size depends upon the measurement used, with CCF time lags and mean values of the radius and time lag showing the largest differences for the LCO 2015 dataset.  
Given the $r_{\rm BLR}-L_{\rm AGN}$ relation \citep{bentz09a, bentz13}, we might expect Arp 151 to have brightened by up to a factor of 4 in the AGN continuum during the 2011 and 2015 campaigns.  

To test whether the AGN continuum flux changed between the three datasets, we remeasured the AGN continuum {\it V}-band light curves with a uniform procedure.  To ensure that the same level of host galaxy flux is included in all three light curves, we used a uniform photometric aperture size of 4\arcsec\ and the same comparison stars for all three campaigns.  We also restricted our analysis to the highest-quality data from each campaign, including Tenagra Observatory (2008), WMO (2011), and LCOGT (2015).  Photometry measurements were made using the automated aperture photometry code described by \citet{pei14}.
The {\it V}-band light curve mean values are $15.5$, $15.5$, and $15.6$~mag for the LAMP 2008, LAMP 2011, and LCO 2015 light curves, respectively.  
Comparing the {\it V}-band light curve mean values provides a lower limit to the AGN variability due to a significant contribution from the host galaxy flux estimated to be $47$\% for LAMP 2008 \citep{walsh09}.

Before including a host galaxy correction, we can first compare the variations in the mean {\it V}-band magnitude to the variability within each reverberation mapping campaign.  Comparing the mean {\it V}-band magnitudes, the standard deviation of the three measurements is 0.057 mag, with a spread between the maximum and minimum value of 0.133 mag.  Calculating these same values for the variability within each dataset, we find a standard deviation (spread between maximum and minimum value) for each light curve of 0.073 (0.322), 0.073 (0.236), and 0.084 (0.339)~mag for LAMP 2008, LAMP 2011, and LCO 2015, respectively.  
This shows that the small changes in the mean {\it V}-band magnitude are less than the variability within each reverberation mapping campaign.
Clearly the BLR size is changing significantly more than expected from the global $r_{\rm BLR}-L_{\rm AGN}$ relation for all AGN \citep{bentz13} given the relatively constant {\it V}-band luminosity.

While it is possible that long-term variations in AGN luminosity or photons at higher energy may be responsible for the larger-than-expected BLR size in 2011 and 2015, there may also be a more complicated $r_{\rm BLR}-L_{\rm AGN}$ relationship for individual AGN.  This last scenario is illustrated by recent reverberation mapping campaigns for NGC 5548, wherein the AGN diverged from its native $r_{\rm BLR}-L_{\rm AGN}$ relation to first increase in AGN luminosity without a corresponding increase in BLR size and then increase in BLR size while actually decreasing in AGN luminosity \citep{pei17}.  Even when NGC 5548 is following its native $r_{\rm BLR}-L_{\rm AGN}$ relation, it is steeper than the global relation and has a scatter of $\sim 0.1$~dex \citep[][]{kilerci15}.  These results suggest that the behavior seen in Arp 151 may not be so unusual.  Only further monitoring will be able to clarify whether its current behavior in $r_{\rm BLR}-L_{\rm AGN}$ space is anomalous or whether it always changes significantly in BLR size at a fixed AGN luminosity.

We can also investigate whether properties of the BLR geometry change with the BLR size.
The minimum radius stays constant to within the statistical uncertainties of $0.2 - 0.5$ light days.  However, the shape of the radial profile does change between the datasets in a manner that appears independent from the size of the BLR, both in terms of the radial distribution shape parameter ($\beta$) and standard deviation ($\sigma_{r}$).  In comparison, other parameters of the BLR geometry, such as the inclination and opening angles, are consistently inferred for all three datasets, while large uncertainties in the inferred asymmetry parameters prevent us from constraining their evolution.  

The BLR dynamics do not appear dependent on the radial BLR size, consistently preferring a majority of inflowing orbits.  However, the LAMP 2011 dataset does prefer a larger fraction of near-circular elliptical orbits, with the peak of the inferred posterior centered on values of $f_{\rm ellip} \sim 0.3$ instead of near $f_{\rm ellip} = 0$.  While this difference is not very large, it is consistent with changes in the velocity-resolved time lag measurements for the three datasets, wherein the LAMP 2008 and LCO 2015 datasets show clear asymmetry in time lag measurements across the \Hb\ line \citep{bentz09, valenti15} while the LAMP 2011 dataset shows more symmetric lag measurements (Barth et al., in preparation).

\subsection{Comparison with MEMEcho Response Functions}
Another way to constrain the properties of the BLR is to recover the broad emission-line response function without assuming a specific BLR model, as discussed in Section~\ref{sect_intro}.  The response functions obtained from regularized linear inversion  \citep{krolik95, done96, skielboe15}  or MEMEcho \citep{horne91, horne94} can then be compared qualitatively to the transfer functions from BLR modeling analysis.  This comparison provides another critical test of the BLR modeling approach because it can show whether the BLR model is flexible enough to produce the response functions found in the data.  In this section we compare transfer functions from BLR modeling with response functions from MEMEcho for the three \Hb\ datasets for Arp 151.  

Full details of the MEMEcho code are described by \citet{horne91} and \citet{horne94}, with previous MEMEcho results for multiple optical broad emission lines for the LAMP 2008 dataset given by \citet{bentz10}.  The datasets used in the MEMEcho analysis were the same as those used for BLR modeling, including reanalysis of the LAMP 2008 dataset with the improved spectral decomposition described in Section~\ref{sect_data}.  
 Note that while BLR modeling fits a {\it linear} echo model
(Eqn.~\ref{eq_2}),
photoionized line emission is in general
a non-linear function of the continuum \citep[e.g.][]{korista04}.
By adopting a linear echo model, values of
$\Psi(v,\tau)$ from BLR modeling can be interpreted
as some mix of the mean and marginal line response.
In contrast, MEMEcho uses a tangent approximation to
the non-linear response, thus fitting a {\it linearised} echo model,
\begin{equation}
        L(\lambda,t) = L_0(\lambda)
        + \int_0^{\tau_{\rm max}} \Psi(\lambda,\tau)
        \left[ C(t-\tau) - C_0 \right]\, d\tau
\ .
\end{equation}
Here $C_0$ is an arbitrary continuum level,
$L_0(\lambda)$ is the line emission corresponding to $C_0$,
and $\Psi(\lambda,\tau)$ is the marginal response,
i.e. the change in line emission per small change in the continuum.
The MEMEcho fit then uses maximum entropy regularization to
find the ``smoothest positive functions''
$C(t)$, $L(\lambda)$, and $\Psi(\lambda,\tau)$
that fit the data at different $\chi^2/N$ levels,
where $N$ is the number of data.
At high $\chi^2/N$ an overly-smooth model under-fits the data
while at low $\chi^2/N$ an overly-noisy model over-fits the data.
A suitable trade-off between these extremes is chosen by eye
to represent the best compromise.
Note that BLR modeling uses MCMC methods to fully sample the joint
posterior probability distribution of its model parameters,
while MEMEcho explores a 1-parameter family of best-fit models,
with uncertainty estimates requiring Monte-Carlo methods.

\begin{figure*}[h!]
\begin{center}
\includegraphics[scale=0.55]{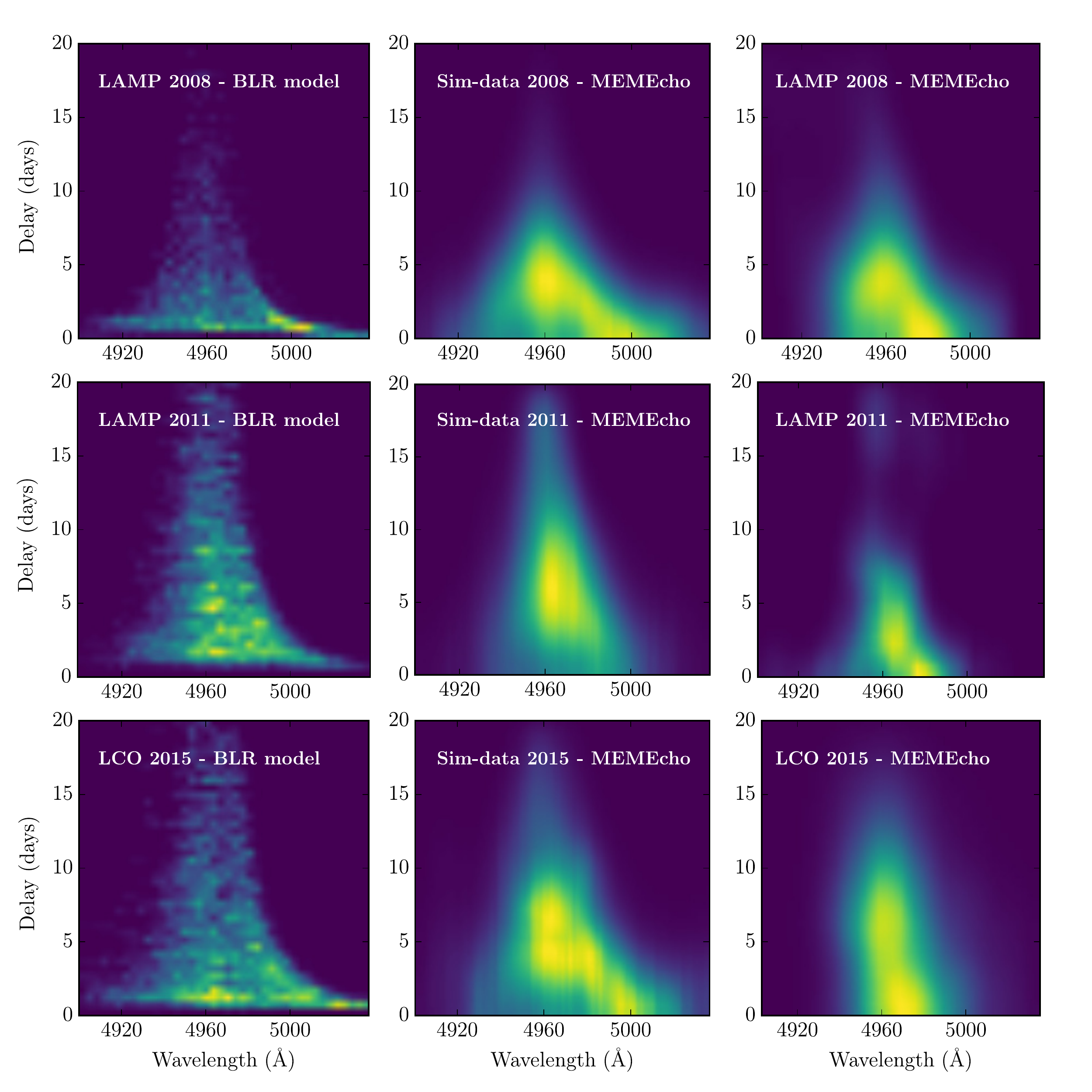}  % left, bottom, right, top
\caption{Comparison of transfer functions from BLR modeling and response functions from MEMEcho.  The top, middle, and bottom rows show results for LAMP 2008, LAMP 2011, and LCO 2015, respectively.  The left column shows a representative transfer function from the posterior PDF from BLR modeling for each dataset.  The middle column shows the MEMEcho response function for a simulated dataset generated from the BLR modeling posterior sample shown in the left column for each dataset.  The right column shows the MEMEcho response function for the data.  Light yellow indicates the most emission in the transfer (response) functions, while dark blue indicates the least, with the absolute scales of emission (response) being relative.  \label{fig_tf}}
\end{center}
\end{figure*}

The transfer functions from BLR modeling and the response functions from MEMEcho are shown in Figure~\ref{fig_tf} as the first and third columns, respectively.  For each Arp~151 dataset, we show in Column 1 a BLR modeling transfer function created from a posterior sample that is chosen such that the transfer function shape is representative of the range shown by many posterior samples.  The second column of Figure~\ref{fig_tf} shows the MEMEcho response function for a simulated dataset created from the posterior sample BLR model shown in the first column.   
Comparing the first and second columns shows that MEMEcho smooths the BLR modeling transfer functions significantly.  This is not surprising, since MEMEcho tries to find the smoothest response function that still fits the data, while the BLR modeling approach allows for any sharp features in the transfer function that can be made using the BLR model, such as the sharp emission feature in the red wing.  This suggests that a better comparison between BLR modeling and MEMEcho can be made by comparing the MEMEcho response functions for the simulated data and the real data (second and third columns).  

However, there are two points to note when making this comparison. First, since the posterior PDFs of the BLR model parameters mostly overlap for the three Arp 151 datasets, as shown in Figure~\ref{fig_final}, the three transfer functions shown in Figure~\ref{fig_tf} can also reasonably be interpreted as showing the range in transfer function shape for any of the individual Arp 151 datasets, with the exception of the difference in average time lag that could cause the transfer function shape to be shifted vertically and compressed horizontally to follow the virial envelope.  This does not mean that all three transfer functions shown are equally likely for all three datasets; instead, the LAMP 2008 and LCO 2015 datasets have a higher fraction of posterior samples with transfer functions showing strong red-wing asymmetry, while the LAMP 2011 dataset has a higher fraction of posterior samples showing a more symmetric transfer function.  These differences in transfer function asymmetry are due to the larger probability of the BLR having a higher fraction of point particles in near-circular elliptical orbits for LAMP 2011.  The second point to note when comparing the second and third columns of Figure~\ref{fig_tf} is that we do not have uncertainty estimates for the MEMEcho response functions, so we cannot make a quantitative comparison.  

From a qualitative perspective, the MEMEcho response functions tend to look similar for the simulated and real datasets, with two main regions of high response: at longer time delays in the middle of the line profile ($\sim 4960$ \AA) and at smaller time delays in the red wing of the line profile ($\sim 4970-5000$ \AA).  In addition, the simulated and real datasets for LAMP 2008 have very similar MEMEcho response functions, suggesting that the BLR model used in this analysis is flexible enough to capture the large-scale response function behavior of the data.  
Overall, these results suggest that BLR modeling and MEMEcho analysis are providing consistent information about the BLR transfer function and response function, respectively.  

%%%%%%%%%%%%%%%%%%%%%%%%%%%%%%%%%%%%%%
%%%%%%%%%%%%%%%%%%%%%%%%%%%%%%%%%%%%%%
%%%%%%%%%%%%%%%%%%%%%%%%%%%%%%%%%%%%%%
%%%%%%%%%%%%%%%%%%%%%%%%%%%%%%%%%%%%%%

\section{Summary}  \label{sect_summary}
We have analyzed three \Hb\ reverberation mapping datasets for Arp 151 taken over seven years using a geometric and dynamical model for the BLR.  By comparing multiple datasets for the same AGN we are able to probe the systematic uncertainties in the inferred \Hb\ BLR structure and look for evolution of the geometry or dynamics on the orbital time.  Our main results are as follows:
\begin{enumerate}
\item The inferred BH mass ranges from $\log_{10}($\mbh$)=$\,\lampaMbh\ to \lampbMbh\ for LAMP 2008 and LAMP 2011, respectively, with a standard deviation in the three measurements of 0.13 dex. Since the individual BH masses agree to within the statistical uncertainties, we calculate the combined inference on the BH mass from all three datasets of $\log_{10}($\mbh$)=$\totMbh, which is the most precise BH mass measurement from BLR modeling to date.  
\item The size of the BLR grows by a factor of $\sim 2$ between 2008 and 2011, although the minimum radius stays the same over all seven years.  The shape of the radial profile of emission and the standard deviation of the radial profile do show small changes for each dataset, although the changes are not correlated with the BLR size.  
\item The inclination angle and opening angle are consistently inferred ($\theta_o=$\,\totThetao\ deg and $\theta_i=$\,\totThetai\ deg), despite the change in size of the BLR. 
\item Each dataset constrains the BLR geometry asymmetry parameters to different degrees.   While the direction of emission back towards the central ionizing source is only constrained by the LAMP 2008 dataset, all three datasets prefer an opaque disk mid-plane, such that a transparent mid-plane is ruled out in a joint inference.  There is also preference for more emission from the faces of the disk for the LAMP 2008 and 2011 datasets.  
\item The BLR dynamics are consistently inferred to be dominated by mostly-radial inflowing orbits, with the LAMP 2011 dataset showing a higher contribution from near-circular elliptical orbits.  These differences are consistent with velocity-resolved time lag analysis.
\item We try to include the systematic uncertainty from spectral decomposition in the statistical uncertainties above by marginalizing over results including and excluding the red wing of \Hb.  Spectral decomposition to isolate the \Hb\ line is sensitive to the choice of \feii\ template and the presence of \hei\ in the \Hb\ red wing at a level that is often greater than the spectral flux uncertainties.  The choice of whether to exclude parts of the \Hb\ profile due to contamination also affects the results by increasing the inferred statistical uncertainties by $3-50$\% depending on the portion of the line excluded and the specific dataset.  This suggests that by improving spectral decomposition techniques to marginalize over the inclusion of different spectral components and templates self-consistently, we can significantly reduce BLR modeling uncertainties in the future. 
\item Comparison between BLR modeling and independent MEMEcho analysis suggests that both methods find similar transfer/response function shapes.
\end{enumerate}

Overall, these results show that parameters expected to be constant in time, such as the BH mass and inclination angle of the BLR, are consistently inferred for Arp 151 from completely independent datasets and analysis.  This suggests that the BLR modeling approach implemented here is robust to reproducibility, although there may still be significant systematic uncertainties introduced by our choice of a specific model.  A lack of large changes in other, potentially time-dependent, BLR model parameters for the three datasets suggests that the \Hb\ BLR structure in Arp 151 is fairly constant on the orbital time, with the exception of the radial size of the BLR that is expected to change in response to variability in the AGN continuum luminosity.  

\acknowledgments
We thank the Lick Observatory staff for their exceptional support during our LAMP 2008 and 2011 observing campaigns.  
AP is supported by NASA through Einstein Postdoctoral Fellowship grant number PF5-160141 awarded by the Chandra X-ray Center, which is operated by the Smithsonian Astrophysical Observatory for NASA under contract NAS8-03060.
Research by AJB has been supported by NSF grant AST-1412693.
V.N.B. acknowledges assistance from a National Science Foundation (NSF) Research at Undergraduate Institutions (RUI) grant AST-1312296. Note that findings and
conclusions do not necessarily represent views of the NSF.
KH acknowledges support from STFC grant ST/M001296/1.
SFH acknowledges support from the EU/Horizon 2020 programme via an ERC Starting Grant 677117 and the UK STFC via grant ST/N000870/1.
Research by DJS is supported by NSF grant AST-1412504 and AST-1517649.
TT gratefully acknowledges support from the Packard Foundations through a Packard Research Fellowship and by NSF through grant AST-1412315.
JHW acknowledges support by the National Research Foundation of Korea (NRF) grant funded by the Korean government (No.2017R1A5A1070354)

\bibliographystyle{apj}
\bibliography{references}

\begin{thebibliography}{}
\expandafter\ifx\csname natexlab\endcsname\relax\def\natexlab#1{#1}\fi

\bibitem[{{Barth} {et~al.}(2013){Barth}, {Pancoast}, {Bennert}, {Brewer},
  {Canalizo}, {Filippenko}, {Gates}, {Greene}, {Li}, {Malkan}, {Sand}, {Stern},
  {Treu}, {Woo}, {Assef}, {Bae}, {Buehler}, {Cenko}, {Clubb}, {Cooper},
  {Diamond-Stanic}, {H{\"o}nig}, {Joner}, {Laney}, {Lazarova}, {Nierenberg},
  {Silverman}, {Tollerud}, \& {Walsh}}]{barth13}
{Barth}, A.~J., {Pancoast}, A., {Bennert}, V.~N., {et~al.} 2013, \apj, 769, 128

\bibitem[{{Barth} {et~al.}(2015){Barth}, {Bennert}, {Canalizo}, {Filippenko},
  {Gates}, {Greene}, {Li}, {Malkan}, {Pancoast}, {Sand}, {Stern}, {Treu},
  {Woo}, {Assef}, {Bae}, {Brewer}, {Cenko}, {Clubb}, {Cooper},
  {Diamond-Stanic}, {Hiner}, {H{\"o}nig}, {Hsiao}, {Kandrashoff}, {Lazarova},
  {Nierenberg}, {Rex}, {Silverman}, {Tollerud}, \& {Walsh}}]{barth15}
{Barth}, A.~J., {Bennert}, V.~N., {Canalizo}, G., {et~al.} 2015, \apjs, 217, 26

\bibitem[{{Batiste} {et~al.}(2017){Batiste}, {Bentz}, {Raimundo},
  {Vestergaard}, \& {Onken}}]{batiste17}
{Batiste}, M., {Bentz}, M.~C., {Raimundo}, S.~I., {Vestergaard}, M., \&
  {Onken}, C.~A. 2017, \apjl, 838, L10

\bibitem[{{Bentz} \& {Katz}(2015)}]{bentz15}
{Bentz}, M.~C., \& {Katz}, S. 2015, \pasp, 127, 67

\bibitem[{{Bentz} {et~al.}(2009{\natexlab{a}}){Bentz}, {Peterson}, {Netzer},
  {Pogge}, \& {Vestergaard}}]{bentz09a}
{Bentz}, M.~C., {Peterson}, B.~M., {Netzer}, H., {Pogge}, R.~W., \&
  {Vestergaard}, M. 2009{\natexlab{a}}, \apj, 697, 160

\bibitem[{{Bentz} {et~al.}(2009{\natexlab{b}}){Bentz}, {Walsh}, {Barth},
  {Baliber}, {Bennert}, {Canalizo}, {Filippenko}, {Ganeshalingam}, {Gates},
  {Greene}, {Hidas}, {Hiner}, {Lee}, {Li}, {Malkan}, {Minezaki}, {Sakata},
  {Serduke}, {Silverman}, {Steele}, {Stern}, {Street}, {Thornton}, {Treu},
  {Wang}, {Woo}, \& {Yoshii}}]{bentz09}
{Bentz}, M.~C., {Walsh}, J.~L., {Barth}, A.~J., {et~al.} 2009{\natexlab{b}},
  \apj, 705, 199

\bibitem[{{Bentz} {et~al.}(2010){Bentz}, {Walsh}, {Barth}, {Yoshii}, {Woo},
  {Wang}, {Treu}, {Thornton}, {Street}, {Steele}, {Silverman}, {Serduke},
  {Sakata}, {Minezaki}, {Malkan}, {Li}, {Lee}, {Hiner}, {Hidas}, {Greene},
  {Gates}, {Ganeshalingam}, {Filippenko}, {Canalizo}, {Bennert}, \&
  {Baliber}}]{bentz10}
---. 2010, \apj, 716, 993

\bibitem[{{Bentz} {et~al.}(2013){Bentz}, {Denney}, {Grier}, {Barth},
  {Peterson}, {Vestergaard}, {Bennert}, {Canalizo}, {De Rosa}, {Filippenko},
  {Gates}, {Greene}, {Li}, {Malkan}, {Pogge}, {Stern}, {Treu}, \&
  {Woo}}]{bentz13}
{Bentz}, M.~C., {Denney}, K.~D., {Grier}, C.~J., {et~al.} 2013, \apj, 767, 149

\bibitem[{{Blandford} \& {McKee}(1982)}]{blandford82}
{Blandford}, R.~D., \& {McKee}, C.~F. 1982, \apj, 255, 419

\bibitem[{{Boroson} \& {Green}(1992)}]{boroson92}
{Boroson}, T.~A., \& {Green}, R.~F. 1992, \apjs, 80, 109

\bibitem[{{Bottorff} {et~al.}(1997){Bottorff}, {Korista}, {Shlosman}, \&
  {Blandford}}]{bottorff97}
{Bottorff}, M., {Korista}, K.~T., {Shlosman}, I., \& {Blandford}, R.~D. 1997,
  \apj, 479, 200

\bibitem[{{Brewer} {et~al.}(2011){Brewer}, {P{\'a}rtay}, \&
  {Cs{\'a}nyi}}]{brewer11}
{Brewer}, B.~J., {P{\'a}rtay}, L.~B., \& {Cs{\'a}nyi}, G. 2011, Statistics and
  Computing, 21, 649, astrophysics Source Code Library

\bibitem[{{Brown} {et~al.}(2013){Brown}, {Baliber}, {Bianco}, {Bowman},
  {Burleson}, {Conway}, {Crellin}, {Depagne}, {De Vera}, {Dilday}, {Dragomir},
  {Dubberley}, {Eastman}, {Elphick}, {Falarski}, {Foale}, {Ford}, {Fulton},
  {Garza}, {Gomez}, {Graham}, {Greene}, {Haldeman}, {Hawkins}, {Haworth},
  {Haynes}, {Hidas}, {Hjelstrom}, {Howell}, {Hygelund}, {Lister}, {Lobdill},
  {Martinez}, {Mullins}, {Norbury}, {Parrent}, {Paulson}, {Petry}, {Pickles},
  {Posner}, {Rosing}, {Ross}, {Sand}, {Saunders}, {Shobbrook}, {Shporer},
  {Street}, {Thomas}, {Tsapras}, {Tufts}, {Valenti}, {Vander Horst}, {Walker},
  {White}, \& {Willis}}]{brown13}
{Brown}, T.~M., {Baliber}, N., {Bianco}, F.~B., {et~al.} 2013, \pasp, 125, 1031

\bibitem[{{Clavel} {et~al.}(1991){Clavel}, {Reichert}, {Alloin}, {Crenshaw},
  {Kriss}, {Krolik}, {Malkan}, {Netzer}, {Peterson}, {Wamsteker}, {Altamore},
  {Baribaud}, {Barr}, {Beck}, {Binette}, {Bromage}, {Brosch}, {Diaz},
  {Filippenko}, {Fricke}, {Gaskell}, {Giommi}, {Glass}, {Gondhalekar},
  {Hackney}, {Halpern}, {Hutter}, {Joersaeter}, {Kinney}, {Kollatschny},
  {Koratkar}, {Korista}, {Laor}, {Lasota}, {Leibowitz}, {Maoz}, {Martin},
  {Mazeh}, {Meurs}, {Nair}, {O'Brien}, {Pelat}, {Perez}, {Perola}, {Ptak},
  {Rodriguez-Pascual}, {Rosenblatt}, {Sadun}, {Santos-Lleo}, {Shaw}, {Smith},
  {Stirpe}, {Stoner}, {Sun}, {Ulrich}, {van Groningen}, \& {Zheng}}]{clavel91}
{Clavel}, J., {Reichert}, G.~A., {Alloin}, D., {et~al.} 1991, \apj, 366, 64

\bibitem[{{Collin} {et~al.}(2006){Collin}, {Kawaguchi}, {Peterson}, \&
  {Vestergaard}}]{collin06}
{Collin}, S., {Kawaguchi}, T., {Peterson}, B.~M., \& {Vestergaard}, M. 2006,
  \aap, 456, 75

\bibitem[{{De Rosa} {et~al.}(2015){De Rosa}, {Peterson}, {Ely}, {Kriss},
  {Crenshaw}, {Horne}, {Korista}, {Netzer}, {Pogge}, {Ar{\'e}valo}, {Barth},
  {Bentz}, {Brandt}, {Breeveld}, {Brewer}, {Dalla Bont{\`a}}, {De
  Lorenzo-C{\'a}ceres}, {Denney}, {Dietrich}, {Edelson}, {Evans}, {Fausnaugh},
  {Gehrels}, {Gelbord}, {Goad}, {Grier}, {Grupe}, {Hall}, {Kaastra}, {Kelly},
  {Kennea}, {Kochanek}, {Lira}, {Mathur}, {McHardy}, {Nousek}, {Pancoast},
  {Papadakis}, {Pei}, {Schimoia}, {Siegel}, {Starkey}, {Treu}, {Uttley},
  {Vaughan}, {Vestergaard}, {Villforth}, {Yan}, {Young}, \& {Zu}}]{derosa15}
{De Rosa}, G., {Peterson}, B.~M., {Ely}, J., {et~al.} 2015, \apj, 806, 128

\bibitem[{{Denney} {et~al.}(2009){Denney}, {Peterson}, {Pogge}, {Adair},
  {Atlee}, {Au-Yong}, {Bentz}, {Bird}, {Brokofsky}, {Chisholm}, {Comins},
  {Dietrich}, {Doroshenko}, {Eastman}, {Efimov}, {Ewald}, {Ferbey}, {Gaskell},
  {Hedrick}, {Jackson}, {Klimanov}, {Klimek}, {Kruse}, {Lad{\'e}route}, {Lamb},
  {Leighly}, {Minezaki}, {Nazarov}, {Onken}, {Petersen}, {Peterson},
  {Poindexter}, {Sakata}, {Schlesinger}, {Sergeev}, {Skolski}, {Stieglitz},
  {Tobin}, {Unterborn}, {Vestergaard}, {Watkins}, {Watson}, \&
  {Yoshii}}]{denney09c}
{Denney}, K.~D., {Peterson}, B.~M., {Pogge}, R.~W., {et~al.} 2009, \apjl, 704,
  L80

\bibitem[{{Denney} {et~al.}(2010){Denney}, {Peterson}, {Pogge}, {Adair},
  {Atlee}, {Au-Yong}, {Bentz}, {Bird}, {Brokofsky}, {Chisholm}, {Comins},
  {Dietrich}, {Doroshenko}, {Eastman}, {Efimov}, {Ewald}, {Ferbey}, {Gaskell},
  {Hedrick}, {Jackson}, {Klimanov}, {Klimek}, {Kruse}, {Lad{\'e}route}, {Lamb},
  {Leighly}, {Minezaki}, {Nazarov}, {Onken}, {Petersen}, {Peterson},
  {Poindexter}, {Sakata}, {Schlesinger}, {Sergeev}, {Skolski}, {Stieglitz},
  {Tobin}, {Unterborn}, {Vestergaard}, {Watkins}, {Watson}, \&
  {Yoshii}}]{denney10}
---. 2010, \apj, 721, 715

\bibitem[{{Done} \& {Krolik}(1996)}]{done96}
{Done}, C., \& {Krolik}, J.~H. 1996, \apj, 463, 144

\bibitem[{{Du} {et~al.}(2016){Du}, {Lu}, {Hu}, {Qiu}, {Li}, {Huang}, {Wang},
  {Bai}, {Bian}, {Yuan}, {Ho}, {Wang}, \& {SEAMBH Collaboration}}]{du16}
{Du}, P., {Lu}, K.-X., {Hu}, C., {et~al.} 2016, \apj, 820, 27

\bibitem[{{Edelson} {et~al.}(2015){Edelson}, {Gelbord}, {Horne}, {McHardy},
  {Peterson}, {Ar{\'e}valo}, {Breeveld}, {De Rosa}, {Evans}, {Goad}, {Kriss},
  {Brandt}, {Gehrels}, {Grupe}, {Kennea}, {Kochanek}, {Nousek}, {Papadakis},
  {Siegel}, {Starkey}, {Uttley}, {Vaughan}, {Young}, {Barth}, {Bentz},
  {Brewer}, {Crenshaw}, {Dalla Bont{\`a}}, {De Lorenzo-C{\'a}ceres}, {Denney},
  {Dietrich}, {Ely}, {Fausnaugh}, {Grier}, {Hall}, {Kaastra}, {Kelly},
  {Korista}, {Lira}, {Mathur}, {Netzer}, {Pancoast}, {Pei}, {Pogge},
  {Schimoia}, {Treu}, {Vestergaard}, {Villforth}, {Yan}, \& {Zu}}]{edelson15}
{Edelson}, R., {Gelbord}, J.~M., {Horne}, K., {et~al.} 2015, \apj, 806, 129

\bibitem[{{Elvis}(2000)}]{elvis00}
{Elvis}, M. 2000, \apj, 545, 63

\bibitem[{{Emmering} {et~al.}(1992){Emmering}, {Blandford}, \&
  {Shlosman}}]{emmering92}
{Emmering}, R.~T., {Blandford}, R.~D., \& {Shlosman}, I. 1992, \apj, 385, 460

\bibitem[{{Fausnaugh} {et~al.}(2016){Fausnaugh}, {Denney}, {Barth}, {Bentz},
  {Bottorff}, {Carini}, {Croxall}, {De Rosa}, {Goad}, {Horne}, {Joner},
  {Kaspi}, {Kim}, {Klimanov}, {Kochanek}, {Leonard}, {Netzer}, {Peterson},
  {Schn{\"u}lle}, {Sergeev}, {Vestergaard}, {Zheng}, {Zu}, {Anderson},
  {Ar{\'e}valo}, {Bazhaw}, {Borman}, {Boroson}, {Brandt}, {Breeveld}, {Brewer},
  {Cackett}, {Crenshaw}, {Dalla Bont{\`a}}, {De Lorenzo-C{\'a}ceres},
  {Dietrich}, {Edelson}, {Efimova}, {Ely}, {Evans}, {Filippenko}, {Flatland},
  {Gehrels}, {Geier}, {Gelbord}, {Gonzalez}, {Gorjian}, {Grier}, {Grupe},
  {Hall}, {Hicks}, {Horenstein}, {Hutchison}, {Im}, {Jensen}, {Jones},
  {Kaastra}, {Kelly}, {Kennea}, {Kim}, {Korista}, {Kriss}, {Lee}, {Lira},
  {MacInnis}, {Manne-Nicholas}, {Mathur}, {McHardy}, {Montouri}, {Musso},
  {Nazarov}, {Norris}, {Nousek}, {Okhmat}, {Pancoast}, {Papadakis}, {Parks},
  {Pei}, {Pogge}, {Pott}, {Rafter}, {Rix}, {Saylor}, {Schimoia}, {Siegel},
  {Spencer}, {Starkey}, {Sung}, {Teems}, {Treu}, {Turner}, {Uttley},
  {Villforth}, {Weiss}, {Woo}, {Yan}, \& {Young}}]{fausnaugh16}
{Fausnaugh}, M.~M., {Denney}, K.~D., {Barth}, A.~J., {et~al.} 2016, \apj, 821,
  56

\bibitem[{{Fausnaugh} {et~al.}(2017){Fausnaugh}, {Grier}, {Bentz}, {Denney},
  {De Rosa}, {Peterson}, {Kochanek}, {Pogge}, {Adams}, {Barth}, {Beatty},
  {Bhattacharjee}, {Borman}, {Boroson}, {Bottorff}, {Brown}, {Brown},
  {Brotherton}, {Coker}, {Crawford}, {Croxall}, {Eftekharzadeh}, {Eracleous},
  {Joner}, {Henderson}, {Holoien}, {Horne}, {Hutchison}, {Kaspi}, {Kim},
  {King}, {Li}, {Lochhaas}, {Ma}, {MacInnis}, {Manne-Nicholas}, {Mason},
  {Montuori}, {Mosquera}, {Mudd}, {Musso}, {Nazarov}, {Nguyen}, {Okhmat},
  {Onken}, {Ou-Yang}, {Pancoast}, {Pei}, {Penny}, {Poleski}, {Rafter},
  {Romero-Colmenero}, {Runnoe}, {Sand}, {Schimoia}, {Sergeev}, {Shappee},
  {Simonian}, {Somers}, {Spencer}, {Starkey}, {Stevens}, {Tayar}, {Treu},
  {Valenti}, {Van Saders}, {Villanueva}, {Villforth}, {Weiss}, {Winkler}, \&
  {Zhu}}]{fausnaugh17}
{Fausnaugh}, M.~M., {Grier}, C.~J., {Bentz}, M.~C., {et~al.} 2017, \apj, 840,
  97

\bibitem[{{Filippenko} {et~al.}(2001){Filippenko}, {Li}, {Treffers}, \&
  {Modjaz}}]{filippenko01}
{Filippenko}, A.~V., {Li}, W.~D., {Treffers}, R.~R., \& {Modjaz}, M. 2001, in
  Astronomical Society of the Pacific Conference Series, Vol. 246, IAU Colloq.
  183: Small Telescope Astronomy on Global Scales, ed. B.~{Paczynski}, W.-P.
  {Chen}, \& C.~{Lemme}, 121

\bibitem[{{Goad} \& {Korista}(2015)}]{goad15}
{Goad}, M.~R., \& {Korista}, K.~T. 2015, \mnras, 453, 3662

\bibitem[{{Graham} {et~al.}(2011){Graham}, {Onken}, {Athanassoula}, \&
  {Combes}}]{graham11}
{Graham}, A.~W., {Onken}, C.~A., {Athanassoula}, E., \& {Combes}, F. 2011,
  \mnras, 412, 2211

\bibitem[{{Greene} {et~al.}(2010){Greene}, {Peng}, {Kim}, {Kuo}, {Braatz},
  {Impellizzeri}, {Condon}, {Lo}, {Henkel}, \& {Reid}}]{greene10b}
{Greene}, J.~E., {Peng}, C.~Y., {Kim}, M., {et~al.} 2010, \apj, 721, 26

\bibitem[{{Grier} {et~al.}(2017){Grier}, {Pancoast}, {Barth}, {Fausnaugh},
  {Brewer}, {Treu}, \& {Peterson}}]{grier17}
{Grier}, C.~J., {Pancoast}, A., {Barth}, A.~J., {et~al.} 2017, \apj, 849, 146

\bibitem[{{Grier} {et~al.}(2012){Grier}, {Peterson}, {Pogge}, {Denney},
  {Bentz}, {Martini}, {Sergeev}, {Kaspi}, {Minezaki}, {Zu}, {Kochanek},
  {Siverd}, {Shappee}, {Stanek}, {Araya Salvo}, {Beatty}, {Bird}, {Bord},
  {Borman}, {Che}, {Chen}, {Cohen}, {Dietrich}, {Doroshenko}, {Drake},
  {Efimov}, {Free}, {Ginsburg}, {Henderson}, {King}, {Koshida}, {Mogren},
  {Molina}, {Mosquera}, {Nazarov}, {Okhmat}, {Pejcha}, {Rafter}, {Shields},
  {Skowron}, {Szczygiel}, {Valluri}, \& {van Saders}}]{grier12}
{Grier}, C.~J., {Peterson}, B.~M., {Pogge}, R.~W., {et~al.} 2012, \apj, 755, 60

\bibitem[{{Grier} {et~al.}(2013{\natexlab{a}}){Grier}, {Martini}, {Watson},
  {Peterson}, {Bentz}, {Dasyra}, {Dietrich}, {Ferrarese}, {Pogge}, \&
  {Zu}}]{grier13b}
{Grier}, C.~J., {Martini}, P., {Watson}, L.~C., {et~al.} 2013{\natexlab{a}},
  \apj, 773, 90

\bibitem[{{Grier} {et~al.}(2013{\natexlab{b}}){Grier}, {Peterson}, {Horne},
  {Bentz}, {Pogge}, {Denney}, {De Rosa}, {Martini}, {Kochanek}, {Zu},
  {Shappee}, {Siverd}, {Beatty}, {Sergeev}, {Kaspi}, {Araya Salvo}, {Bird},
  {Bord}, {Borman}, {Che}, {Chen}, {Cohen}, {Dietrich}, {Doroshenko}, {Efimov},
  {Free}, {Ginsburg}, {Henderson}, {King}, {Mogren}, {Molina}, {Mosquera},
  {Nazarov}, {Okhmat}, {Pejcha}, {Rafter}, {Shields}, {Skowron}, {Szczygiel},
  {Valluri}, \& {van Saders}}]{grier13a}
{Grier}, C.~J., {Peterson}, B.~M., {Horne}, K., {et~al.} 2013{\natexlab{b}},
  \apj, 764, 47

\bibitem[{{Horne}(1994)}]{horne94}
{Horne}, K. 1994, in Astronomical Society of the Pacific Conference Series,
  Vol.~69, Reverberation Mapping of the Broad-Line Region in Active Galactic
  Nuclei, ed. P.~M. {Gondhalekar}, K.~{Horne}, \& B.~M. {Peterson}, 23--25

\bibitem[{{Horne} {et~al.}(1991){Horne}, {Welsh}, \& {Peterson}}]{horne91}
{Horne}, K., {Welsh}, W.~F., \& {Peterson}, B.~M. 1991, \apjl, 367, L5

\bibitem[{{Kaspi} {et~al.}(2007){Kaspi}, {Brandt}, {Maoz}, {Netzer},
  {Schneider}, \& {Shemmer}}]{kaspi07}
{Kaspi}, S., {Brandt}, W.~N., {Maoz}, D., {et~al.} 2007, \apj, 659, 997

\bibitem[{{Kaspi} {et~al.}(2005){Kaspi}, {Maoz}, {Netzer}, {Peterson},
  {Vestergaard}, \& {Jannuzi}}]{kaspi05}
{Kaspi}, S., {Maoz}, D., {Netzer}, H., {et~al.} 2005, \apj, 629, 61

\bibitem[{{Kaspi} {et~al.}(2000){Kaspi}, {Smith}, {Netzer}, {Maoz}, {Jannuzi},
  \& {Giveon}}]{kaspi00}
{Kaspi}, S., {Smith}, P.~S., {Netzer}, H., {et~al.} 2000, \apj, 533, 631

\bibitem[{{Kilerci Eser} {et~al.}(2015){Kilerci Eser}, {Vestergaard},
  {Peterson}, {Denney}, \& {Bentz}}]{kilerci15}
{Kilerci Eser}, E., {Vestergaard}, M., {Peterson}, B.~M., {Denney}, K.~D., \&
  {Bentz}, M.~C. 2015, \apj, 801, 8

\bibitem[{{King} {et~al.}(2015){King}, {Martini}, {Davis}, {Denney},
  {Kochanek}, {Peterson}, {Skielboe}, {Vestergaard}, {Huff}, {Watson},
  {Banerji}, {McMahon}, {Sharp}, \& {Lidman}}]{king15}
{King}, A.~L., {Martini}, P., {Davis}, T.~M., {et~al.} 2015, \mnras, 453, 1701

\bibitem[{{Korista} \& {Goad}(2004)}]{korista04}
{Korista}, K.~T., \& {Goad}, M.~R. 2004, \apj, 606, 749

\bibitem[{{Kova{\v c}evi{\'c}} {et~al.}(2010){Kova{\v c}evi{\'c}},
  {Popovi{\'c}}, \& {Dimitrijevi{\'c}}}]{kovacevic10}
{Kova{\v c}evi{\'c}}, J., {Popovi{\'c}}, L.~{\v C}., \& {Dimitrijevi{\'c}},
  M.~S. 2010, \apjs, 189, 15

\bibitem[{{Krolik} \& {Done}(1995)}]{krolik95}
{Krolik}, J.~H., \& {Done}, C. 1995, \apj, 440, 166

\bibitem[{{Li} {et~al.}(2013){Li}, {Wang}, {Ho}, {Du}, \& {Bai}}]{li13}
{Li}, Y.-R., {Wang}, J.-M., {Ho}, L.~C., {Du}, P., \& {Bai}, J.-M. 2013, \apj,
  779, 110

\bibitem[{{Lu} {et~al.}(2016){Lu}, {Du}, {Hu}, {Li}, {Zhang}, {Wang}, {Huang},
  {Bi}, {Bai}, {Ho}, \& {Wang}}]{lu16}
{Lu}, K.-X., {Du}, P., {Hu}, C., {et~al.} 2016, \apj, 827, 118

\bibitem[{{MacLeod} {et~al.}(2010){MacLeod}, {Ivezi{\'c}}, {Kochanek},
  {Koz{\l}owski}, {Kelly}, {Bullock}, {Kimball}, {Sesar}, {Westman}, {Brooks},
  {Gibson}, {Becker}, \& {de Vries}}]{macleod10}
{MacLeod}, C.~L., {Ivezi{\'c}}, {\v Z}., {Kochanek}, C.~S., {et~al.} 2010,
  \apj, 721, 1014

\bibitem[{{McConnell} \& {Ma}(2013)}]{mcconnell13}
{McConnell}, N.~J., \& {Ma}, C.-P. 2013, \apj, 764, 184

\bibitem[{{Miller} \& {Stone}(1993)}]{miller93}
{Miller}, J.~S., \& {Stone}, R.~P.~S. 1993, Lick Obs. Tech. Rep. (Santa Cruz,
  CA: Lick Observatory)

\bibitem[{{Murray} \& {Chiang}(1997)}]{murray97}
{Murray}, N., \& {Chiang}, J. 1997, \apj, 474, 91

\bibitem[{{Onken} {et~al.}(2004){Onken}, {Ferrarese}, {Merritt}, {Peterson},
  {Pogge}, {Vestergaard}, \& {Wandel}}]{onken04}
{Onken}, C.~A., {Ferrarese}, L., {Merritt}, D., {et~al.} 2004, \apj, 615, 645

\bibitem[{{Pancoast} {et~al.}(2014{\natexlab{a}}){Pancoast}, {Brewer}, \&
  {Treu}}]{pancoast14a}
{Pancoast}, A., {Brewer}, B.~J., \& {Treu}, T. 2014{\natexlab{a}}, \mnras, 445,
  3055

\bibitem[{{Pancoast} {et~al.}(2014{\natexlab{b}}){Pancoast}, {Brewer}, {Treu},
  {Park}, {Barth}, {Bentz}, \& {Woo}}]{pancoast14b}
{Pancoast}, A., {Brewer}, B.~J., {Treu}, T., {et~al.} 2014{\natexlab{b}},
  \mnras, 445, 3073

\bibitem[{{Pancoast} {et~al.}(2012){Pancoast}, {Brewer}, {Treu}, {Barth},
  {Bennert}, {Canalizo}, {Filippenko}, {Gates}, {Greene}, {Li}, {Malkan},
  {Sand}, {Stern}, {Woo}, {Assef}, {Bae}, {Buehler}, {Cenko}, {Clubb},
  {Cooper}, {Diamond-Stanic}, {Hiner}, {H{\"o}nig}, {Joner}, {Kandrashoff},
  {Laney}, {Lazarova}, {Nierenberg}, {Park}, {Silverman}, {Son}, {Sonnenfeld},
  {Thorman}, {Tollerud}, {Walsh}, \& {Walters}}]{pancoast12}
---. 2012, \apj, 754, 49

\bibitem[{{Park} {et~al.}(2017){Park}, {Barth}, {Woo}, {Malkan}, {Treu},
  {Bennert}, {Assef}, \& {Pancoast}}]{park17}
{Park}, D., {Barth}, A.~J., {Woo}, J.-H., {et~al.} 2017, \apj, 839, 93

\bibitem[{{Park} {et~al.}(2012){Park}, {Kelly}, {Woo}, \& {Treu}}]{park12b}
{Park}, D., {Kelly}, B.~C., {Woo}, J.-H., \& {Treu}, T. 2012, \apjs, 203, 6

\bibitem[{{Pei} {et~al.}(2014){Pei}, {Barth}, {Aldering}, {Briley}, {Carroll},
  {Carson}, {Cenko}, {Clubb}, {Cohen}, {Cucchiara}, {Desjardins}, {Edelson},
  {Fang}, {Fedrow}, {Filippenko}, {Fox}, {Furniss}, {Gates}, {Gregg},
  {Gustafson}, {Horst}, {Joner}, {Kelly}, {Lacy}, {Laney}, {Leonard}, {Li},
  {Malkan}, {Margon}, {Neeleman}, {Nguyen}, {Prochaska}, {Ross}, {Sand},
  {Searcy}, {Shivvers}, {Silverman}, {Smith}, {Suzuki}, {Smith}, {Tytler},
  {Werk}, \& {Worseck}}]{pei14}
{Pei}, L., {Barth}, A.~J., {Aldering}, G.~S., {et~al.} 2014, \apj, 795, 38

\bibitem[{{Pei} {et~al.}(2017){Pei}, {Fausnaugh}, {Barth}, {Peterson}, {Bentz},
  {De Rosa}, {Denney}, {Goad}, {Kochanek}, {Korista}, {Kriss}, {Pogge},
  {Bennert}, {Brotherton}, {Clubb}, {Dalla Bont{\`a}}, {Filippenko}, {Greene},
  {Grier}, {Vestergaard}, {Zheng}, {Adams}, {Beatty}, {Bigley}, {Brown},
  {Brown}, {Canalizo}, {Comerford}, {Coker}, {Corsini}, {Croft}, {Croxall},
  {Deason}, {Eracleous}, {Fox}, {Gates}, {Henderson}, {Holmbeck}, {Holoien},
  {Jensen}, {Johnson}, {Kelly}, {Kim}, {King}, {Lau}, {Li}, {Lochhaas}, {Ma},
  {Manne-Nicholas}, {Mauerhan}, {Malkan}, {McGurk}, {Morelli}, {Mosquera},
  {Mudd}, {Muller Sanchez}, {Nguyen}, {Ochner}, {Ou-Yang}, {Pancoast}, {Penny},
  {Pizzella}, {Poleski}, {Runnoe}, {Scott}, {Schimoia}, {Shappee}, {Shivvers},
  {Simonian}, {Siviero}, {Somers}, {Stevens}, {Strauss}, {Tayar}, {Tejos},
  {Treu}, {Van Saders}, {Vican}, {Villanueva}, {Yuk}, {Zakamska}, {Zhu},
  {Anderson}, {Ar{\'e}valo}, {Bazhaw}, {Bisogni}, {Borman}, {Bottorff},
  {Brandt}, {Breeveld}, {Cackett}, {Carini}, {Crenshaw}, {De
  Lorenzo-C{\'a}ceres}, {Dietrich}, {Edelson}, {Efimova}, {Ely}, {Evans},
  {Ferland}, {Flatland}, {Gehrels}, {Geier}, {Gelbord}, {Grupe}, {Gupta},
  {Hall}, {Hicks}, {Horenstein}, {Horne}, {Hutchison}, {Im}, {Joner}, {Jones},
  {Kaastra}, {Kaspi}, {Kelly}, {Kennea}, {Kim}, {Kim}, {Klimanov}, {Lee},
  {Leonard}, {Lira}, {MacInnis}, {Mathur}, {McHardy}, {Montouri}, {Musso},
  {Nazarov}, {Netzer}, {Norris}, {Nousek}, {Okhmat}, {Papadakis}, {Parks},
  {Pott}, {Rafter}, {Rix}, {Saylor}, {Schn{\"u}lle}, {Sergeev}, {Siegel},
  {Skielboe}, {Spencer}, {Starkey}, {Sung}, {Teems}, {Turner}, {Uttley},
  {Villforth}, {Weiss}, {Woo}, {Yan}, {Young}, \& {Zu}}]{pei17}
{Pei}, L., {Fausnaugh}, M.~M., {Barth}, A.~J., {et~al.} 2017, \apj, 837, 131

\bibitem[{{Peterson}(1993)}]{peterson93}
{Peterson}, B.~M. 1993, \pasp, 105, 247

\bibitem[{{Peterson}(2014)}]{peterson14}
---. 2014, \ssr, 183, 253

\bibitem[{{Peterson} \& {Wandel}(1999)}]{peterson99}
{Peterson}, B.~M., \& {Wandel}, A. 1999, \apjl, 521, L95

\bibitem[{{Peterson} \& {Wandel}(2000)}]{peterson00}
---. 2000, \apjl, 540, L13

\bibitem[{{Peterson} {et~al.}(2004){Peterson}, {Ferrarese}, {Gilbert}, {Kaspi},
  {Malkan}, {Maoz}, {Merritt}, {Netzer}, {Onken}, {Pogge}, {Vestergaard}, \&
  {Wandel}}]{peterson04}
{Peterson}, B.~M., {Ferrarese}, L., {Gilbert}, K.~M., {et~al.} 2004, \apj, 613,
  682

\bibitem[{{Proga} {et~al.}(2000){Proga}, {Stone}, \& {Kallman}}]{proga00}
{Proga}, D., {Stone}, J.~M., \& {Kallman}, T.~R. 2000, \apj, 543, 686

\bibitem[{{Reichert} {et~al.}(1994){Reichert}, {Rodriguez-Pascual}, {Alloin},
  {Clavel}, {Crenshaw}, {Kriss}, {Krolik}, {Malkan}, {Netzer}, {Peterson},
  {Wamsteker}, {Altamore}, {Altieri}, {Anderson}, {Blackwell}, {Boisson},
  {Brosch}, {Carone}, {Dietrich}, {England}, {Evans}, {Filippenko}, {Gaskell},
  {Goad}, {Gondhalekar}, {Horne}, {Kazanas}, {Kollatschny}, {Koratkar},
  {Korista}, {MacAlpine}, {Maoz}, {Mazeh}, {McCollum}, {Miller}, {Mendes de
  Oliveira}, {O'Brien}, {Pastoriza}, {Pelat}, {Perez}, {Perola}, {Pogge},
  {Ptak}, {Recondo-Gonzalez}, {Rodriguez-Espinosa}, {Rosenblatt}, {Sadun},
  {Santos-Lleo}, {Shields}, {Shrader}, {Shull}, {Simkin}, {Sitko}, {Snijders},
  {Sparke}, {Stirpe}, {Stoner}, {Storchi-Bergmann}, {Sun}, {Wang}, {Welsh},
  {White}, {Winge}, \& {Zheng}}]{reichert94}
{Reichert}, G.~A., {Rodriguez-Pascual}, P.~M., {Alloin}, D., {et~al.} 1994,
  \apj, 425, 582

\bibitem[{{Shen}(2013)}]{shen13}
{Shen}, Y. 2013, Bulletin of the Astronomical Society of India, 41, 61

\bibitem[{{Shen} {et~al.}(2011){Shen}, {Richards}, {Strauss}, {Hall},
  {Schneider}, {Snedden}, {Bizyaev}, {Brewington}, {Malanushenko},
  {Malanushenko}, {Oravetz}, {Pan}, \& {Simmons}}]{shen11}
{Shen}, Y., {Richards}, G.~T., {Strauss}, M.~A., {et~al.} 2011, \apjs, 194, 45

\bibitem[{{Shen} {et~al.}(2015){Shen}, {Brandt}, {Dawson}, {Hall}, {McGreer},
  {Anderson}, {Chen}, {Denney}, {Eftekharzadeh}, {Fan}, {Gao}, {Green},
  {Greene}, {Ho}, {Horne}, {Jiang}, {Kelly}, {Kinemuchi}, {Kochanek},
  {P{\^a}ris}, {Peters}, {Peterson}, {Petitjean}, {Ponder}, {Richards},
  {Schneider}, {Seth}, {Smith}, {Strauss}, {Tao}, {Trump}, {Wood-Vasey}, {Zu},
  {Eisenstein}, {Pan}, {Bizyaev}, {Malanushenko}, {Malanushenko}, \&
  {Oravetz}}]{shen15a}
{Shen}, Y., {Brandt}, W.~N., {Dawson}, K.~S., {et~al.} 2015, \apjs, 216, 4

\bibitem[{{Shen} {et~al.}(2016){Shen}, {Horne}, {Grier}, {Peterson}, {Denney},
  {Trump}, {Sun}, {Brandt}, {Kochanek}, {Dawson}, {Green}, {Greene}, {Hall},
  {Ho}, {Jiang}, {Kinemuchi}, {McGreer}, {Petitjean}, {Richards}, {Schneider},
  {Strauss}, {Tao}, {Wood-Vasey}, {Zu}, {Pan}, {Bizyaev}, {Ge}, {Oravetz}, \&
  {Simmons}}]{shen16}
{Shen}, Y., {Horne}, K., {Grier}, C.~J., {et~al.} 2016, \apj, 818, 30

\bibitem[{{Skielboe} {et~al.}(2015){Skielboe}, {Pancoast}, {Treu}, {Park},
  {Barth}, \& {Bentz}}]{skielboe15}
{Skielboe}, A., {Pancoast}, A., {Treu}, T., {et~al.} 2015, \mnras, 454, 144

\bibitem[{{Valenti} {et~al.}(2015){Valenti}, {Sand}, {Barth}, {Horne}, {Treu},
  {Raganit}, {Boroson}, {Crawford}, {Pancoast}, {Pei}, {Romero-Colmenero},
  {Villforth}, \& {Winkler}}]{valenti15}
{Valenti}, S., {Sand}, D.~J., {Barth}, A.~J., {et~al.} 2015, \apjl, 813, L36

\bibitem[{{V{\'e}ron-Cetty} {et~al.}(2004){V{\'e}ron-Cetty}, {Joly}, \&
  {V{\'e}ron}}]{veron-cetty04}
{V{\'e}ron-Cetty}, M.-P., {Joly}, M., \& {V{\'e}ron}, P. 2004, \aap, 417, 515

\bibitem[{{Vestergaard} \& {Peterson}(2006)}]{vestergaard06}
{Vestergaard}, M., \& {Peterson}, B.~M. 2006, \apj, 641, 689

\bibitem[{{Walsh} {et~al.}(2009){Walsh}, {Minezaki}, {Bentz}, {Barth},
  {Baliber}, {Li}, {Stern}, {Bennert}, {Brown}, {Canalizo}, {Filippenko},
  {Gates}, {Greene}, {Malkan}, {Sakata}, {Street}, {Treu}, {Woo}, \&
  {Yoshii}}]{walsh09}
{Walsh}, J.~L., {Minezaki}, T., {Bentz}, M.~C., {et~al.} 2009, \apjs, 185, 156

\bibitem[{{Whittle}(1992)}]{whittle92}
{Whittle}, M. 1992, \apjs, 79, 49

\bibitem[{{Woo} {et~al.}(2013){Woo}, {Schulze}, {Park}, {Kang}, {Kim}, \&
  {Riechers}}]{woo13}
{Woo}, J.-H., {Schulze}, A., {Park}, D., {et~al.} 2013, \apj, 772, 49

\bibitem[{{Woo} {et~al.}(2015){Woo}, {Yoon}, {Park}, {Park}, \& {Kim}}]{woo15}
{Woo}, J.-H., {Yoon}, Y., {Park}, S., {Park}, D., \& {Kim}, S.~C. 2015, \apj,
  801, 38

\bibitem[{{Woo} {et~al.}(2010){Woo}, {Treu}, {Barth}, {Wright}, {Walsh},
  {Bentz}, {Martini}, {Bennert}, {Canalizo}, {Filippenko}, {Gates}, {Greene},
  {Li}, {Malkan}, {Stern}, \& {Minezaki}}]{woo10}
{Woo}, J.-H., {Treu}, T., {Barth}, A.~J., {et~al.} 2010, \apj, 716, 269

\end{thebibliography}

\end{document}